\documentclass[draft]{agujournal2019}
\usepackage{url} 
\usepackage{lineno}
\usepackage[inline]{trackchanges} 
\usepackage{soul}

\usepackage{xurl}           

\draftfalse
\begin{document}

\title{
Cold pools, Breezes, and Monsoons: Propagating Convection over New Guinea}

\authors{Mingyue Tang\affil{1,2,4}, Jimy Dudhia\affil{2}, Changhai Liu\affil{3}, Giuseppe Torri\affil{1}}

\affiliation{1}{Department of Atmospheric Sciences, University of Hawai`i at M\={a}noa, Honolulu, HI, USA}
\affiliation{2}{Mesoscale and Microscale Meteorology Laboratory, NSF National Center for Atmospheric Research, Boulder, CO, USA}
\affiliation{3}{Research Applications Laboratory, NSF National Center for Atmospheric Research, Boulder, CO, USA}
\affiliation{4}{Present address: Earth System Physics, International Centre for Theoretical Physics, Trieste, TS, Italy}

\correspondingauthor{Mingyue Tang}{mtang@ictp.it}


\begin{keypoints}
\item New Guinea’s diurnal offshore-propagating convection has two modes: ridge-to-coast and over-ocean, separated by a clear jump.
\item Moist patches over warm ocean are observed not only at cold pool fronts but also along land breeze fronts.
\item Hybrid land breeze and cold pools promote offshore convective propagation.
\end{keypoints}

%
%


\begin{abstract}
The diurnal cycle of precipitation near New Guinea involves intricate land-ocean-atmosphere interactions, posing substantial challenges for tropical weather and climate simulations. Using over two decades of GPM satellite observations and convection-permitting WRF simulations, this study examines the physical mechanisms governing the pronounced offshore propagation of diurnal convection over New Guinea. We identify two distinct convective propagation modes: (1) a “ridge-to-coast” mode originated over elevated terrain and migrating toward the coastline, and (2) an “over-ocean” mode initiated near the coast, separated by a spatial gap of approximately 100 km. Our findings highlight the critical role of multi-scale thermally driven flow in shaping boundary-layer dynamics over warm ocean waters. Specifically, the afternoon sea-breeze front advects cooler air onshore, stabilizing the lower atmosphere and interrupting the continuous propagation of the first mode. At night, the hybrid land breeze, strengthened by cold pools, generates offshore moist patches that facilitate the convective regeneration and propagation of the second mode. These offshore convective systems interact with monsoonal background winds, sustaining precipitation well beyond 200–600 km from the coast. Sensitivity experiments indicate that even a modest increase in sea surface temperature can enhance convective intensity and extend offshore propagation. These results shed light on the mechanisms that enable diurnal offshore convection to persist overnight and propagate far from the coastline, highlighting the importance of moist-boundary-layer density currents and offering insights for improving precipitation forecasts and global model performance over the Maritime Continent.
\end{abstract}

\section*{Plain Language Summary}
Near the island of New Guinea, thunderstorms and rain clouds often start over mountains in the afternoon and then move toward the coast. Surprisingly, many of these storms seem to “jump” across the shoreline and regenerate over the ocean at night, sometimes traveling hundreds of kilometers offshore. This study uses satellite rainfall data and specialized computer simulations to uncover how different “density currents” in the lower atmosphere—including daily winds flowing between land and sea (“land-sea breezes”) and cooler air spreading out from rainstorms (“cold pools”)—work together to push storms far into the ocean. During the daytime, cooler air from the sea breeze flows onshore until evening, creating a gap in the migrating storm. After sunset, the wind direction reverses, and a land breeze forms; if it merges with pockets of cool, rainy air, it can carry enough moisture out to sea to form new storms. Ocean waters that are relatively warm provide extra energy and moisture, further supporting overnight rainstorms far from land. Our findings help explain why some coastal areas experience heavy rain at unusual times of day and offer clues for improving weather and climate models in tropical regions, where accurate rainfall forecasts are critical.

%
%
\section{Introduction}

The diurnal cycle of convection is one of the leading modes of climate variability in the tropics. One of the places where this phenomenon is particularly important is the Maritime Continent, partly due to the so-called \textit{barrier effect} that inhibits eastward Madden-Julian Oscillation (MJO) propagation, potentially through diurnal precipitation over islands and adjacent seas \cite{ling2019MCbarrier}. However, the diurnal cycle over the Maritime Continent is relatively unique among tropical islands \cite{nesbitt2003, peatman2014MC}, strongly influenced by the interplay between thermally and mechanically forced mesoscale dynamics \cite{wang2017island}. The offshore propagation of convection—a process that enables the propagation of rainfall from land to neighboring seas—plays a crucial role in determining both the timing and intensity of coastal precipitation. This phenomenon accounts for roughly 60\% of the rainfall observed near coastlines globally and demonstrates remarkable persistency over the Maritime Continent \cite{fang2022global}. However, uncertainties in representing this process limit our understanding of diurnal convection and pose challenges for accurately simulating tropical precipitation in global climate models \cite{neale2003MC}.

New Guinea, the largest and most topographically prominent island in the Maritime Continent, is distinguished by its towering mountain ranges, which exceed 4 km in elevation. This distinctive geography influences the diurnal cycle of convection and its offshore propagation \cite{zhou2006NGtopoGW, peatman2021ensomjo}. Typically, afternoon precipitation peaks over land with onshore migration of the sea breeze front \cite{zhou2006NGtopoGW, kikuchi2008mc}, and dynamic propagation events extending northeastward from the island \cite{hassim2016diurnal}. Moreover, precipitation patterns throughout New Guinea demonstrate strong seasonality, primarily driven by the monsoons \cite{mcalpine1983NG}. During the boreal winter rainy season \cite{mcalpine1983NG}, cold and dry northeasterly winds from the Northern Hemisphere cross the equator, transforming into westerlies due to the Coriolis effect \cite{wang2006MS}. These boreal-wintertime monsoonal flows interact with the elevated terrain, substantially modulating diurnal precipitation patterns \cite{zhou2006NGtopoGW}. However, the precise mechanisms underlying these processes, including the roles of land-sea breezes, cold pools, monsoons, gravity waves, and their interactions, remain insufficiently understood.

A long-standing unsolved issue concerns the physical mechanism driving the offshore propagation of the diurnal convection. This phenomenon has been attributed either to gravity waves or to density currents. However, additional factors such as background winds \cite{wang2017island, lopez2023, XC2024GWmonsoon}, topography \cite{coppin2019topo}, convective initiation over the ocean \cite{XC2025longdisOP}, the life cycle and organization of mesoscale convective systems (MCSs) \cite{lopez2023}, and the seeder-feeder effect of anvils \cite{mori2004} complicate the dynamics of offshore propagation further.

Density currents, such as cold pools, land breezes, katabatic winds, and mountain breezes, offer near-surface forcing that supports the life cycle and the offshore propagation of convective systems with a range of speeds of approximately 2–8 m s$^{-1}$ \cite{houze1981monsoonOP, mori2004, hassim2016diurnal, vincent2016NGmjo, coppin2019topo, liu2000DC, peatman2021ensomjo, lopez2023, peatman2023, stoddard2024, peatman2025}. It has been recognized that the diurnal cycle of precipitation is most pronounced during periods of minimal large-scale pressure gradient variations \cite{chen2008, hsiao2021}. However, some studies highlight the significance of monsoon-induced surface convergence \cite{houze1981monsoonOP} and the interactions between land breezes and monsoonal flows in enhancing diurnal convective offshore propagation \cite{wu2009monsoonOP}.

Gravity waves can arise from several processes, including daily temperature fluctuations over land and ocean \cite{rotunno1983linear, ruppert2019diurnalGW}, daytime convection occurring over land \cite{zhou2006NGtopoGW, yokoi2017,coppin2019flat, ruppert2020convGW}, and the propagating organized convection \cite{liu2017}. The cooling phases of these waves can destabilize the lower atmosphere and enhance Convective Available Potential Energy (CAPE) ahead of propagating convection, thereby facilitating the upward and outward propagation of energy to help offshore propagation \cite{yang2001, mapes2003GWop,warner2003, hassim2016diurnal, XC2024GWmonsoon}. 

Recent research has argued that gravity waves play a dominant role in tropical offshore propagation, especially in situations where density currents cannot extend beyond 100 km to maintain offshore precipitation \cite{fang2022global}. On the other hand, \citeA{hassim2016diurnal} noted that diurnally organized convection over the northeast coast of New Guinea propagates offshore at approximately 5 m s$^{-1}$, reaching nearly 200 km from the coastline. This propagation speed is considerably lower than the typical speeds exceeding 15 m s$^{-1}$ associated with dry hydrostatic gravity waves, yet it is comparable to the speed characteristics of tropical density currents.
However, the propagation distance far surpasses the traditionally anticipated range for density currents \cite{love2011}. This indicates that density currents, especially in the form of land breezes, may facilitate convection extending farther offshore over warm oceans during nighttime, a mechanism that merits further exploration. 

Thus, two hypotheses involving the interaction and resemblance between land breezes and cold pools may help explain the contrast noted above, which we refer to collectively as the \textit{hybrid-land-breeze hypothesis}:
\begin{enumerate}
   \item Cold pools serve as an additional driving force, deepening and strengthening land breezes generated by nocturnal radiative cooling and hence enhancing their propagation speed and persistence. 
   \item Land breezes propagating over warm oceans develop \textit{moist patches}, areas with anomalously high water vapor, similar to those found in
cold pools \cite{tompkins2001, tang2024}, thus facilitating new convection triggering at their leading edges. The moist patches in cold pools are mainly attributed to rain evaporation in downdrafts and surface moisture fluxes \cite{14vaporring, schlemmer2016, torri2016MP}. It is plausible that similar processes provide an additional source of moisture to the leading edge of land breezes. 
\end{enumerate}

Together, the mechanisms described above can support convective systems and facilitate their offshore propagation generally along the head of hybrid land breezes over long distances and periods of time. Furthermore, these processes are influenced by strong, relatively cooler, denser yet more humid sea breezes \cite{wang2017island,miller2003SB}, as well as cooler and drier cross-equatorial monsoons traveling in the low troposphere. Moreover, we propose that the moist thermodynamics of the atmospheric boundary layer, shaped by interactions between thermally driven flows of various scales, cooling resources, and the warm ocean environment, plays a crucial role in driving and regulating the diurnal cycle and offshore propagation of convection over the island of New Guinea.

This study seeks to clarify the mechanisms underlying the diurnal offshore propagation of convection over New Guinea and the surrounding ocean by addressing the following questions:
 \begin{itemize}
   \item How do multi-scale density currents influence the diurnal offshore propagation of convection over New Guinea?
   \item What factors enable organized convection to propagate such long-distance offshore and persist for such an extended duration over the ocean?
   \item What is the nature of the asymmetry and interaction between observed land breezes and sea breezes, and what role do cold pools play in this relationship and in facilitating propagation?
\end{itemize}

Section \ref{sec:methods} describes the observational datasets, numerical model, and the simulations and sensitivity experiments used for this work; Section \ref{sec:results} presents the results from both observations and simulations. Finally, an interpretation of the results along with a discussion of their implications and the conclusions are presented in sections \ref{sec:discussion} and \ref{sec:conclusions}, respectively.

\section{Methods}
\label{sec:methods}

\subsection{Observational Data} 

To analyze the climatological characteristics of diurnal precipitation and its offshore propagation, we used precipitation data from the Integrated Multi-satellite Retrievals for GPM (IMERG version 6) product, covering 21 years (2001–2021) \cite{huffman2019}. IMERG provides near-global coverage by combining observations from multiple satellite platforms, including data from the Tropical Rainfall Measuring Mission (TRMM) before the launch of GPM in 2014, ensuring a consistent long-term record of rainfall estimates. With a high spatial-temporal resolution of $0.1^\circ$ and 30 minutes, IMERG is particularly suited for detecting mesoscale convective features and capturing the diurnal cycle over both land and ocean. 

We characterized large-scale environmental conditions using ERA5 reanalysis \cite{hersbach2020}, which provides hourly atmospheric fields at 0.25° × 0.25° horizontal resolution. Unless otherwise noted, all time references are given in Western New Guinea local time (UTC+9). Additionally, in the Hovmöller diagrams presented in this manuscript, time increases from bottom to top to facilitate visualization of the temporal evolution.

This study focuses on the month of February, which falls within the summer in the Southern Hemisphere, where New Guinea is situated. The period from January to March constitutes the island’s rainy season, characterized by abundant precipitation. February was selected as the focal time for several reasons: 1) its alignment with prior research for comparison, notably \citeA{hassim2016diurnal}, who examined conditions from February 2010 and examined physical mechanisms; 2) its diurnal offshore propagation of convection over New Guinea is the most pronounced of the year (see Movie S1 in Supporting Information); 3) its strongest northerly component of the cross-equatorial monsoonal flow \cite{wang2006MS}, which \citeA{zhou2006NGtopoGW} posited is a key factor influencing the diurnal cycle of precipitation over New Guinea. For the sake of comparison, August was also analyzed, as it is the month that exhibits the strongest southerly component of the cross-equatorial monsoonal flow.

\subsection{Numerical model and experiment design} 

We conducted numerical simulations using the Weather Research and Forecasting (WRF) Model, version 4.5.2 \cite{skamarock2019WRFV4}, along with the WRF Preprocessing System (WPS) version 4.5 to generate initial and lateral boundary conditions. The model domain was configured on a Mercator projection, covering New Guinea and adjacent equatorial oceanic regions, with a horizontal grid spacing of 2 km (1390 × 1110 grid points shown in Figure \ref{fig:2coors}a). The vertical grid comprised 61 hybrid sigma-pressure levels, extending from the surface to 25 hPa. The integration time step was 12 s, and outputs were saved at hourly intervals.

The selected physics suite included the WSM6 microphysics scheme \cite{hong2006WSM6}, RRTMG radiation schemes \cite{iacono2008} for both shortwave and longwave radiation, and the Yonsei University (YSU) planetary boundary layer scheme coupled with the Monin–Obukhov surface-layer parameterization \cite{hong2006}. Surface processes were represented by the Noah-MP land surface model \cite{niu2011}. Given the 2 km grid spacing, cumulus parameterization was deactivated to allow for explicit convection. Geographical inputs included MODIS land-use data and high-resolution terrain data (30 arc-second) from GMTED2010. The lateral boundary and initial conditions were derived from ERA5 reanalysis.

The control simulation covered 3 days, from 00:00 UTC on 19 February 2010 until 00:00 UTC on 22 February 2010. During this period, a prominent long-range
 offshore propagation event, extending over 600 km, was observed in the northeast side of the island (Figure \ref{fig:Hov.PR-GPM_C1_SST}, right). This feature, not well captured in \citeA{hassim2016diurnal}, appeared only in the positive X channel. The negative X channel showed offshore propagation extending 200–300 km, closely resembling the climatological pattern. While other cases were also simulated, we focus on this event for its representativeness and possible underlying mechanisms.

The winter of 2010 coincided with an El Niño phase, which was associated with slightly cooler SST anomalies northeast of New Guinea (see Figure S1 in Supporting Information). To assess the role of SST in convective offshore propagation, we conducted a perturbed \textit{SST(+0.5 K) Experiment}, in which SSTs were uniformly increased by 0.5 K across the domain as the initial and lower-boundary conditions. The perturbation amplitude of +0.5 K represents a modest but non-negligible warming, allowing us to isolate the thermodynamic contribution of SST to surface fluxes and convective organization without introducing unrealistic boundary forcing. This design highlights whether a warmer ocean surface favors sustained offshore propagation. 

To investigate the impact of cold pools on offshore propagation and their interactions with sea–land breezes, we performed a \textit{Cold Pool Experiment}, in which the evaporative cooling effect was disabled from the WSM6 microphysics scheme while evaporation was retained. To ensure numerical stability under these modified conditions, the integration time step was reduced to 8 seconds.

\subsection{Definitions}
\label{definitions}

We utilize density potential temperature, $\theta_\rho$ \cite{emanuel1994book} to characterize density currents including cold pools, defined as

\begin{equation}
\theta_\rho = \theta \left[ 1 + \left( \frac{R_v}{R_d} - 1 \right) q_v - q_l \right]
\end{equation}

where $\theta$ is the potential temperature. The term $q_v$ represents the water vapor mixing ratio, while $q_l$ is the liquid water mixing ratio. $R_d$ and $R_v$ are the gas constants for dry air and water vapor, respectively.

The buoyancy of an air parcel relative to its environment is given by the reduced gravity:

\begin{equation}
B = g \left( \frac{\theta_{\rho} - \theta_{\rho,e}}{\theta_{\rho,e}} \right)
\end{equation}

where $g$ is the gravitational acceleration, 
$\theta_{\rho}$ is the density potential temperature of the air parcel, 
and $\theta_{\rho,e}$ is the density potential temperature of the environment.

Finally, moist static energy (MSE) is defined as:

\begin{equation}
h = c_p T + gz + L_v q_v.
\end{equation}

In this equation, $c_p$ is the specific heat capacity of dry air at constant pressure, 
$T$ is temperature, $z$ is height, and 
$L_v$ is the latent heat of vaporization.

Here we define a density current as a near-surface flow composed of air whose $\theta_{\rho}$ is lower than that of its surroundings; such air is negatively buoyant and spreads laterally along the surface under gravity through the horizontal pressure gradients it induces. A temperature deficit of just 1 K relative to the environment already represents a pronounced signature of density currents \cite{torri2015, tang2024}, with the full estimation given by Eq. (4) in Section 4.2. Density currents are ubiquitous in the atmosphere, including cold pools and sea-land breezes. These currents can travel vast distances and influence local weather and climate. For instance, air parcels ahead of the gust front, being warmer and moister with higher density potential temperatures than the currents behind, can be lifted to trigger convection.


Unlike other density currents, cold pools originate from precipitation-driven downdrafts. In addition to their negative $\theta_{\rho}$ anomaly, they are characterized by enhanced column-integrated liquid water content below 2 km and the downward transport of low-MSE air, which is further cooled by evaporation and lower boundary-layer MSE. We therefore diagnose cold-pool thermodynamics using a two-and-a-half-variable framework based on $\theta_{\rho}$, MSE, and column-integrated liquid water content below 2 km along the cross-sections (applied in Section 3.3.3). Negative $\theta_{\rho}$ anomaly identifies the dense, negatively buoyant air mass itself, while negative MSE above it helps distinguish downdrafts and cooling from non-precipitating density currents such as land breezes. Liquid water content serves as a complementary structural tracer: when centered near the overlap of negative $\theta_{\rho}$ anomaly and negative MSE anomalies, it indicates old convective cells associated with mature cold pools; when located near the edge of the negative $\theta_{\rho}$ anomaly, it more often marks newly developing convective cells, typically near gust fronts or moist patches with positive MSE anomalies. This diagnostic is used primarily to identify the spatial relationship between old and new convective cells and to track how cold pools connect them. It also provides a framework to assess whether hybrid land breezes exhibit density-current characteristics and include embedded cold-pool components.

Land/sea breezes are commonly regarded as coastal density currents that propagate relative to the coastline, driven by the temperature contrast between land and sea.  A sea breeze refers to the daytime scenario, in which relatively cooler and denser marine air advances inland, while a land breeze denotes the nighttime reversal, in which colder and denser land air moves offshore. These definitions emphasize not only the diurnal wind reversal but also the density-current structure oscillating across the coastline. The sea and land breeze fronts mark the thermodynamic and dynamical boundary between the current head and the surrounding air, essentially acting as gust fronts. Because the near-surface flow within the current body can travel at a speed distinct from that of the leading head or front, the local wind speeds inland and over the ocean vary accordingly during the propagation of the sea-breeze and land-breeze fronts.

\section{Results}
\label{sec:results}
\subsection{Climatology} 


\begin{figure}
 \noindent\includegraphics[width=39pc,angle=0]{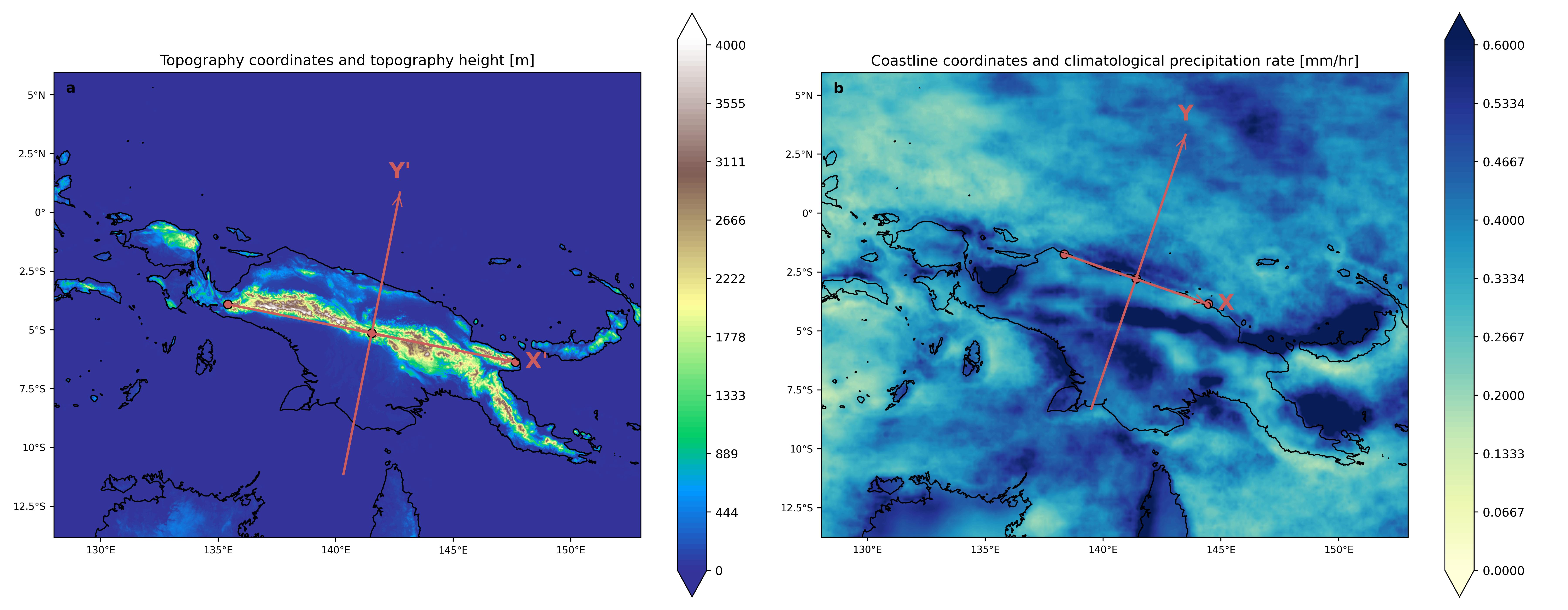}\\
 \caption{Maps showing (a) topography height (m) and (b) climatological precipitation rate (mm hr$^{-1}$) in February using hourly GPM data within the WRF domain. Two coordinate systems are defined for Hovmöller diagrams. The x-axis is aligned with the mountain ridge (X'-Y') (a), and the northeast coastline of New Guinea (X-Y) (b). The y-axis extends offshore in the northeastern direction. The x-axis in the coastline-aligned coordinate system (b) is divided into negative and positive channels to account for the geographically varying influence of cross-equatorial monsoon flows.}
 \label{fig:2coors}
\end{figure}

\begin{figure}
 \noindent\includegraphics[width=39pc,angle=0]{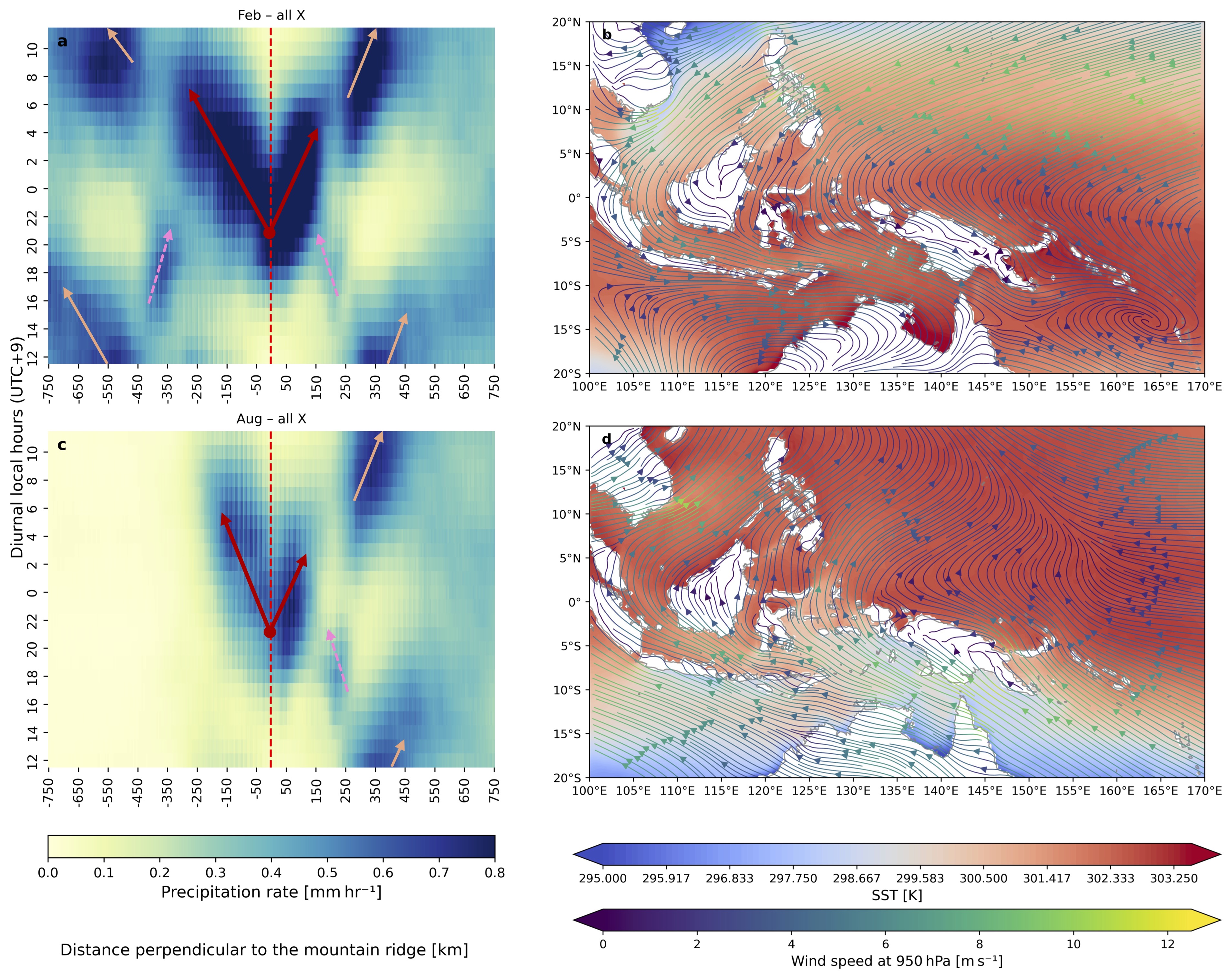}\\
 \caption{February (top) and August (bottom):
(Left) Hovmöller diagrams showing the averaged diurnal cycle of precipitation rate (mm hr$^{-1}$), based on 21 years of GPM data and plotted in a topography-aligned coordinate system  (X'-Y') (Figure \ref{fig:2coors}a). The red dashed line marks the ridge location. Red arrows denote dominant daily offshore-propagating convection originating near the ridge (first mode), while pink dashed arrows indicate weaker onshore-propagating convection at the sea breeze front. Light orange arrows highlight further offshore extension following the decay of the main convective signal.
(Right) Climatological SST (shading in K) and 950 hPa wind streamlines (colored by wind speed in m s$^{-1}$), derived from the 21-year averaged ERA5 dataset.}
 \label{fig:clim.Feb.Aug.mtn}
\end{figure}

The topography of New Guinea, illustrated in Figure \ref{fig:2coors}a, is defined by complex terrain dominated by steep mountain ridges exceeding 4 km in height. This pronounced orography exerts a strong control on regional circulations and substantially influences convection and precipitation. In particular, the central ridges have been shown to play a pivotal role in initiating convection within the diurnal precipitation cycle, effectively splitting rainbands \cite{peatman2021ensomjo} and modulating their subsequent propagation. Previous research has largely concentrated on the northeastern coastline \cite{hassim2016diurnal, peatman2021ensomjo, fang2022global}. Still, the question of whether similar and symmetric offshore propagation occurs along the island’s southwestern side related to the ridge remains largely unexamined. To better analyze the convection emanating from the primary ridges toward both sides symmetrically, we introduce the first orthogonal coordinate system, with the x-axis aligned along the ridge and the positive y-axis oriented toward the northeast. This system is designated as the \textit{topography-aligned coordinate system} (X'-Y'). This coordinate system is shown in Figure \ref{fig:2coors}a, and it will be used for some of the analyses for Figure \ref{fig:clim.Feb.Aug.mtn}.

Figure \ref{fig:2coors}b illustrates the February-averaged precipitation rate for 2001-2021 over New Guinea, the adjacent seas, and surrounding islands based on observational data within the same domain utilized for the subsequent WRF simulations. The rainiest areas in the domain can be roughly divided into three modes: rainfall occurring over the slopes of central mountains; rainfall originating over the ocean near the island's coastline; and rainfall situated over the northern equatorial ocean (142-151°E, 0-6°N), far from the islands.  Two distinct dry gaps separate the rainbands over the mountain slopes from those over the adjacent coastal waters. 

Furthermore, the coastline plays a critical role in modulating convection and precipitation patterns around the tropical islands, primarily due to the marked thermal and moisture contrasts between land and ocean surfaces, as well as the localized circulations they induce. To facilitate the investigation of distinct land-water conditions, associated processes, and transitions across the coastline, a second coordinate system is introduced in Figure \ref{fig:2coors}b, with the x-axis aligned along the nearly linear northeastern coastline and the positive y-axis directed offshore. This system is designated as the \textit{coastline-aligned coordinate system} (X-Y); unless stated otherwise, we will use the coastline-aligned coordinate system in this manuscript, and ``coastline" refers specifically to the northeastern coast. In subsequent analyses, the coordinate axes are divided into \textit{negative and positive X channels} based on the two sides of the Y-axis to explore the geographically varying influence of cross-equatorial monsoons.

The left panels of Figure \ref{fig:clim.Feb.Aug.mtn} depict the diurnal cycle of precipitation rates averaged over February (Figure \ref{fig:clim.Feb.Aug.mtn}a) and August (Figure \ref{fig:clim.Feb.Aug.mtn}c), highlighting the propagation signature along both flanks of the island ridge. The results indicate that the primary diurnal convective initiation begins in the afternoon along the ridge, around 14 LT. After this initial triggering, ridge convection intensifies through the late afternoon and evening, reaching a stronger precipitation signal around 20 LT. The convection then propagates roughly \textit{symmetrically} down both flanks of the ridge toward the coast (\textit{first mode}), covering distances of approximately 130--230 km along both the positive and negative Y' axes. This is followed by a temporary suppression in convective activity, lasting about 2--4 hours, before convection reemerges offshore (\textit{second mode}), with the exception of the southwestern direction during August.

The plots also show the presence of sea breezes, marked by dashed arrows, that become noticeable in the early afternoon and lead to an onshore propagation of convection. Notably, the convergence between the onshore-moving sea breezes and the offshore propagating convection takes place precisely at the point where convection weakens, as described above. After the offshore convective regeneration, precipitation signals extend further over the ocean, surpassing 750 km, although the precipitation rate tapers off beyond approximately 500 km.

Based on these plots, the offshore propagating speed of precipitation originating over the island mountain slopes generally ranges from 6 to 11 m s$^{-1}$ (Figures \ref{fig:clim.Feb.Aug.mtn}, left). This is faster but overall comparable to the commonly observed speeds of tropical density currents. We propose that cold pools, generated by precipitation-induced evaporative cooling and descending along sloped terrain, primarily drive the rapid convective propagation exceeding 8 m s$^{-1}$ over the island interior. This process can be modulated by large-scale background winds and diurnal orographic circulations manifesting as katabatic winds.

The diurnal cycle of convection and its offshore propagation display pronounced seasonal variations. Comparison between Figures \ref{fig:clim.Feb.Aug.mtn}a and \ref{fig:clim.Feb.Aug.mtn}c suggests that the largest differences occur over the southwestern flank of the island (left part of the plots). In August, precipitation rates are lower, and inland convective signals are largely absent in the afternoon, with no subsequent onshore convection developing at the sea-breeze front or farther offshore over the ocean.

Figures \ref{fig:clim.Feb.Aug.mtn}b,d present the low-level large-scale background fields for February and August, respectively, featuring SST, wind streamlines, and wind speeds at 950 hPa. In February, the large-scale circulation is characterized by a robust cross-equatorial monsoon flow transporting cold air from the Northern to the Southern Hemisphere. In August, the wind flow reverses, exhibiting a stronger southerly component. Zonal winds also shift from easterlies to westerlies in the equatorial region, creating a distinctive curvature in the cross-equatorial monsoon flow due to Coriolis forcing. 
Lower SSTs of approximately 298.4 K are observed along the southwestern coast in August, in contrast to the February scenario, where SSTs range from 302 to 303 K. Meanwhile, relatively cool and dry air associated with the cross-equatorial monsoonal flow is advected from the Southern Hemisphere into the lower troposphere between 950 and 700 hPa (see Figures S2 and S3 in Supporting Information). The combination of reduced SST and this drier lower-tropospheric air likely suppresses boundary-layer moistening and convective initiation near the sea-breeze front and over the adjacent ocean in August. This absence is more pronounced on the island’s southern, more off-equatorial flank. Such spatial variability suggests that boundary-layer thermodynamics, particularly SST, may regulate the regeneration and offshore propagation of the second mode. However, despite advection by the background flow (see Movie 1 in Supporting Information), the velocity component of the first mode perpendicular to the ridge shows little seasonal variation.

\begin{figure}
 \noindent\includegraphics[width=39pc,angle=0]{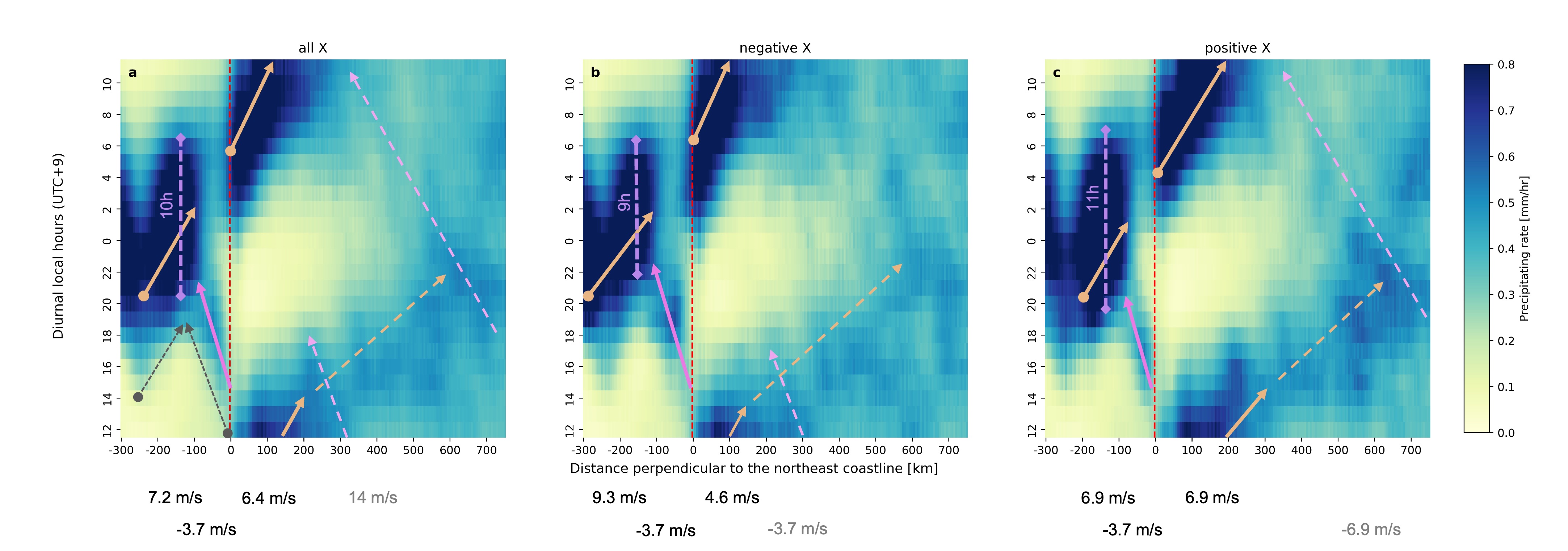}\\
 \caption{Hovmöller diagrams of the averaged diurnal precipitation rate (mm hr$^{-1}$) for February, based on 21 years of GPM data. Panels show results for all x-channels (a), negative x-channels (b), and positive x-channels (c) using the coastline-aligned coordinate system (X-Y). The red dashed line represents the northeast coastline. Orange arrows indicate convective offshore propagation, with circular arrowheads marking the ridge convective core. Pink arrows represent convective onshore propagation. Solid arrows illustrate the movement of deep convection near the coastline (precipitation rates generally exceeding 0.8 mm hr$^{-1}$), while dashed arrows depict the movement of weaker, noisier convection farther offshore with lower precipitation rates. The purple dashed line marks the longest-lasting boundary of deep convection extending from the ridge to the coastline, with its duration (hours) labeled in purple text. Below each diagram, black text indicates the calculated speed of stable near-coast convection, and gray text indicates the speed of unstable offshore convection. Speeds in the top row correspond to offshore propagation, while those in the bottom row represent onshore propagation. Two light-gray dashed guide arrows in (a) indicate the approximate ridge-to-coast distance and the timing relationship between the first mode and the sea-breeze front, together helping infer the gap between the two modes.}
 \label{fig:clim.Hov.coast}
\end{figure}

To more accurately describe the locations of the onshore signal and the regeneration relative to the coastline, Figure \ref{fig:clim.Hov.coast} illustrates the February-averaged diurnal cycle of precipitation rates in the coastline-aligned coordinates (X-Y), illustrating how precipitation propagates across the north and south parts of the northeastern coastline. Over land (negative horizontal axis), deep convection originating from the ridge dissipates at approximately 100 km inland from the coastline. However, the main offshore propagation rainband appears to jump directly to the coastline and regenerate over the ocean (positive horizontal axis), \textit{precisely at the coastline}. This results in a \textit{distinct gap}, characterized by significantly suppressed rainfall,  between the inland and offshore rainbands.

The mechanisms responsible for the formation of the robust gap between the two modes and driving the jump process, as well as their influence on subsequent convective regeneration, are illustrated in Figure \ref{fig:clim.Hov.coast}a, along
 with further quantitative details. The first mode initiates around 2 PM over the ridge or slopes, while weaker convection associated with the sea breeze emerges at 12 PM, 2 hours earlier near the coastline. The two features, approximately 250 km apart, propagate toward each other at speeds of 7.2 m s$^{-1}$ offshore and 3.7 m s$^{-1}$ onshore, respectively, which should converge after 6 hours, at about the halfway point of the distance, 100 km, well matching the distance the first mode and cold air trapped in the sea breezes have traveled by that time. These convergence processes are consistent between the negative and positive X axis (Figure \ref{fig:clim.Hov.coast}b,c), and the convergence zone corresponds to the highest local precipitation rates of the first mode (see Figure S4 in Supporting Information). 

Enhanced by this convergence, the organized precipitation persists for 9 hours in the negative x-channel and 11 hours in the positive x-channel (Figure \ref{fig:clim.Hov.coast}b,c), spanning for nearly half of the diurnal cycle.  The second mode propagates more slowly over the ocean (4.6~m~s$^{-1}$) compared to the offshore speed of the first mode (9.3~m~s$^{-1}$). This difference is likely due to the first mode being accelerated by katabatic winds, terrain slope, and possibly strong cold pools (Figure \ref{fig:clim.Hov.coast}b). However, when convection becomes more organized and persists longer (Figure \ref{fig:clim.Hov.coast}c), the second mode can accelerate, reaching speeds comparable to the first mode (6.9 m s$^{-1}$). Furthermore, the strong convection in the second mode can propagate farther offshore  (by an additional 150 km) and last longer (by 1.5 hours more). Particularly in the positive X channel, the propagation distance reaching 300 km far exceeds the traditionally assumed land breeze influence range within 100 km. This contrast between the positive and negative X channels mainly arises from differences in ridge–coast distance and latitude. The cross-equatorial monsoon brings cool, dry air that accumulates on the equator, tending to suppress the convection in the negative X channel, whereas the positive X channel remains more favorable for a longer-lasting first convective mode and a faster, longer-distance offshore propagation of the second mode (as illustrated in Section \ref{sec:monsoon}).

\subsection{Control run and sensitivity to SST} 

\begin{figure}
 \noindent\includegraphics[width=29pc,angle=0]{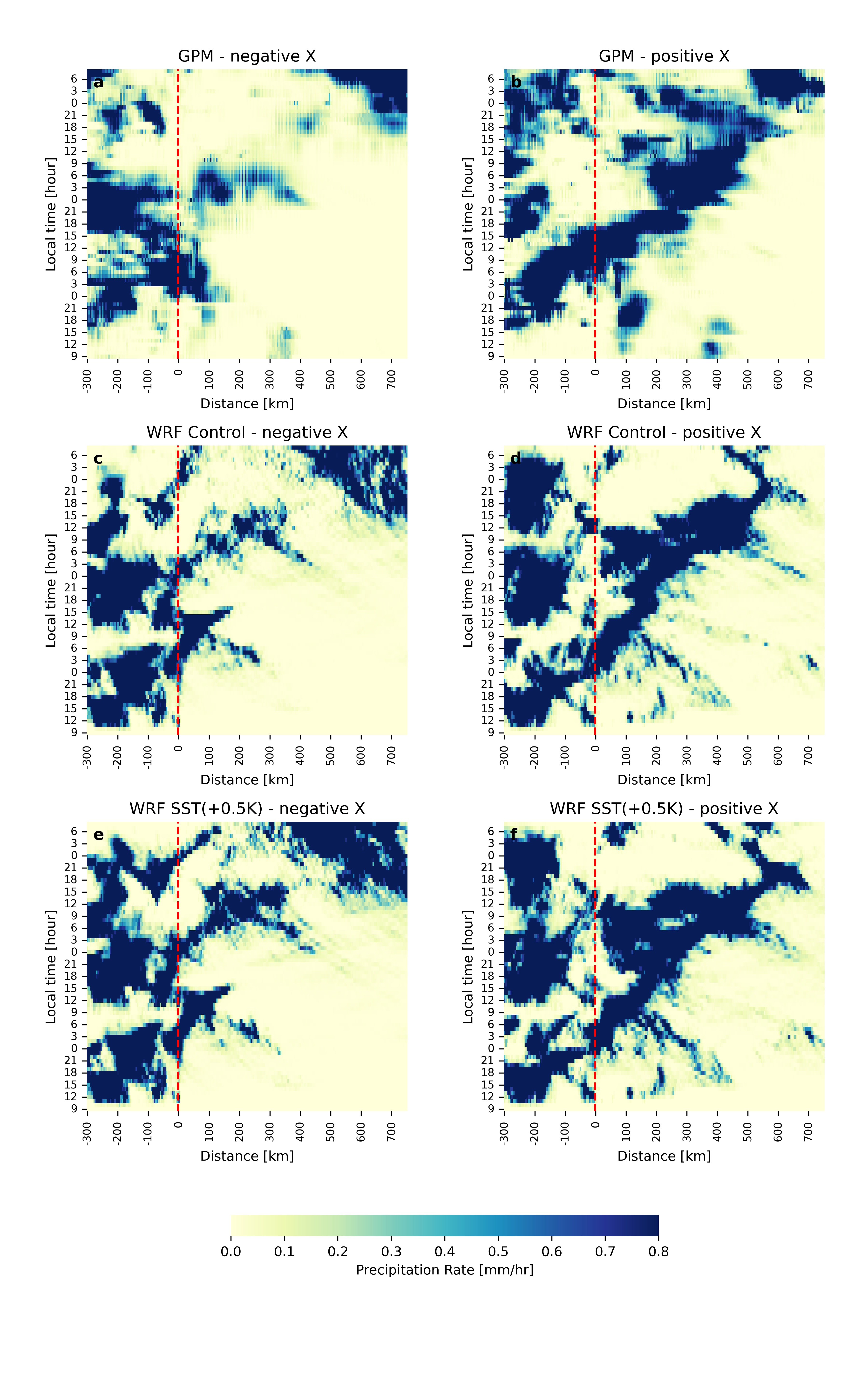}\\
 \caption{Hovmöller diagrams of precipitation rate (mm hr$^{-1}$) from the case study in GPM (top), WRF control run (middle), and WRF SST (+0.5 K) experiment (bottom). Panels show results for negative (left) and positive  (right) X channels, using the coastline-aligned coordinate system (X-Y).}
 \label{fig:Hov.PR-GPM_C1_SST}
\end{figure}

The foregoing analysis of the climatology of observed precipitation reveals intricate diurnal patterns over New Guinea and its surrounding regions, including dual offshore-propagating modes, an apparent suppressed area in between, and propagating convection over the open waters. To investigate the physical mechanisms underlying these diurnal features, convection-permitting simulations of a case study have been conducted, as described in Section 2. The use of high-resolution numerical simulations allows us to complement satellite observations with detailed dynamical and thermodynamical fields, and to explicitly analyze individual cold pools that cannot be isolated in climatological composites.
Figure \ref{fig:Hov.PR-GPM_C1_SST} illustrates the spatiotemporal evolution of precipitation for the same February 2010 case described in Section 2. The control run compares favorably with observations, effectively capturing the extended offshore precipitation event exceeding 600 km in the positive X channel during the second diurnal cycle. The long-distance offshore convective signal that appears on the second to third day of the simulation can be interpreted as the second diurnal mode developing under particularly favorable conditions into a mesoscale convective event. Such events do not occur regularly with the diurnal cycle, but often represent occasions when offshore-propagating convection extends beyond the usual diurnal constraint of 24 hours (section \ref{sec: cold pool regimes}). In the negative X channel, stronger precipitation reaches about 200 km offshore in the first cycle, while weaker precipitation extends to 350 km in the second. The model tends to initiate land convection earlier and produce more organized precipitation than observed, so the observed jump sometimes appears in the simulation as a weakening of the first mode’s precipitation as it continues advancing toward the coast within about 100 km.

In the negative X channel, the effect of orographic precipitation, induced by banded hills ($\sim$1 km in elevation) located approximately 50 km onshore, as shown in Figure \ref{fig:2coors}, is more pronounced in the model. Nevertheless, the approximately 100 km jump between the two offshore propagation convection modes remains a distinct and stable characteristic in the WRF simulation, aligning with the GPM climatology depicted in Figure \ref{fig:clim.Hov.coast}.

To examine the thermodynamic sensitivity (section 2.2), we imposed a uniform 0.5 K increase in SST. Although this perturbation is much smaller than the typical temperature variations experienced on land within a diurnal cycle, we observe that it leads to a slight enhancement of offshore precipitation over the ocean in the negative X channel (Figure \ref{fig:Hov.PR-GPM_C1_SST}c,e) during the second diurnal cycle. Focusing on this period, we find that precipitation triggered by the onshore sea breeze in the afternoon also intensifies in the positive X channel.

\begin{figure}[t]
 \noindent\includegraphics[width=29pc,angle=0]{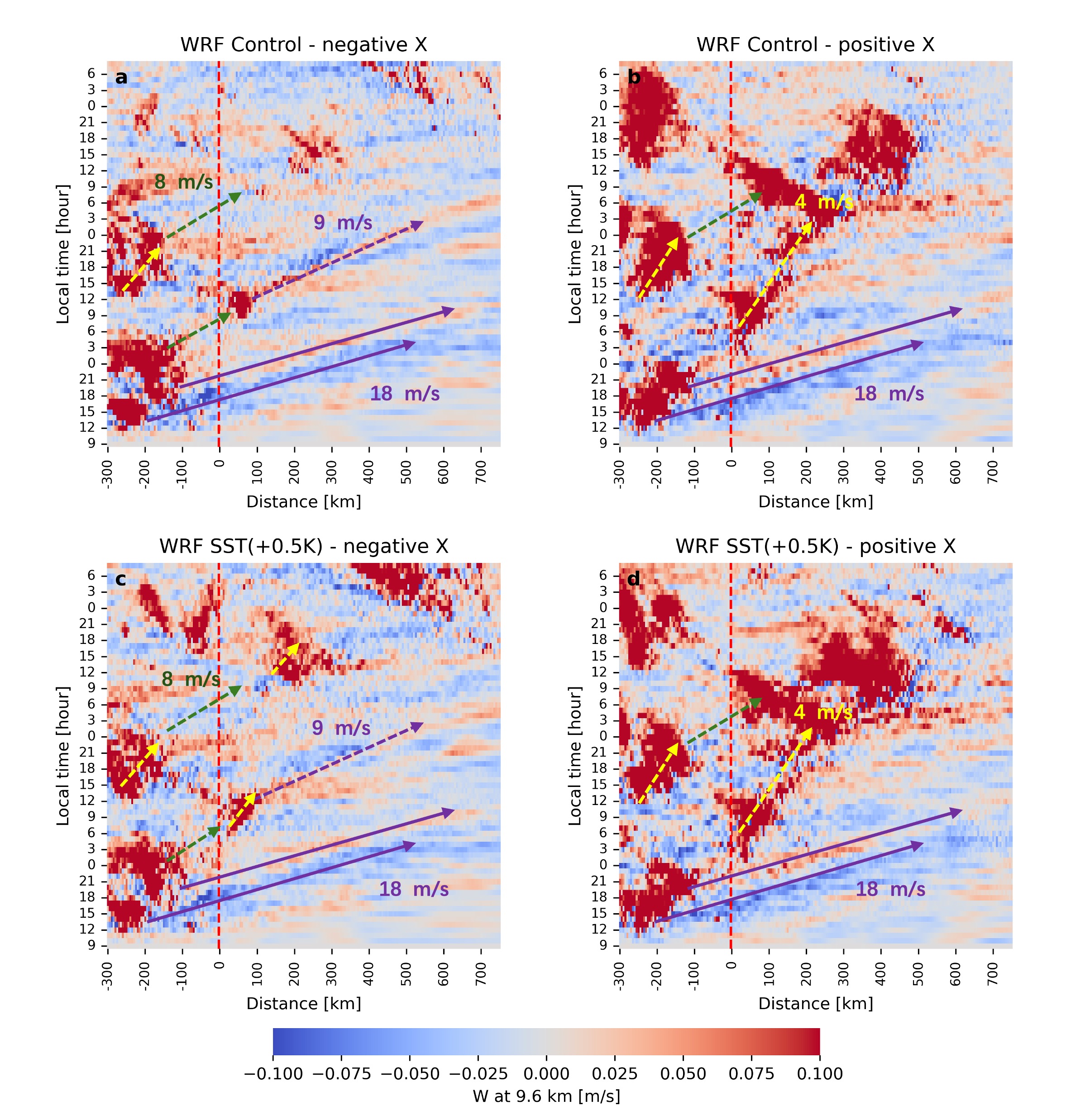}\\
 \caption{Hovmöller diagrams of vertical velocity at an altitude of 9.7 km (free atmosphere) from the case study in the control run (top) and the SST experiment (bottom). Panels show results for negative (left) and positive (right) X channels using the coastline-aligned coordinate system. Solid purple arrows represent gravity wave signals triggered by diurnal convection over the island, while dashed purple arrows represent gravity wave signals generated by the “over-ocean” second mode of convection. Dashed yellow arrows indicate the offshore propagation (offshore propagation) of deep convection, and dashed green arrows depict the ``jump" between the first and second modes of convective offshore propagation. The estimated propagation speeds (m s$^{-1}$) of these signals are labeled next to the respective arrows.}
 \label{fig:Hov.W.10km}
\end{figure}

Figure \ref{fig:Hov.W.10km} illustrates the Hovmöller diagram of vertical velocity at an altitude of 9.6 km based on the simulation results. Across both the negative and positive X channels, and under varying SST conditions, pronounced gravity waves propagate offshore from the island at an estimated speed of approximately 18 m s$^{-1}$. Furthermore, the regenerated second mode of convection over the ocean also appears to produce gravity waves that propagate offshore at a speed of around 9 m s$^{-1}$.

The signatures of both the first and second modes of convective precipitation are clearly observed at this altitude, propagating offshore at approximately 4 m s$^{-1}$, much slower than gravity waves. The convective updrafts remain strong above 9.6 km and persist for an extended period, indicating that the precipitation highlighted by the yellow arrows is associated with organized MCSs. Moreover, convection regenerates offshore with a timing consistent with an  8 m s$^{-1}$ fast jump (dashed green arrows), suggesting that although precipitation weakens during this transition, it does not hinder the overall offshore movement of precipitation. 

\begin{figure}
 \noindent\includegraphics[width=29pc,angle=0]{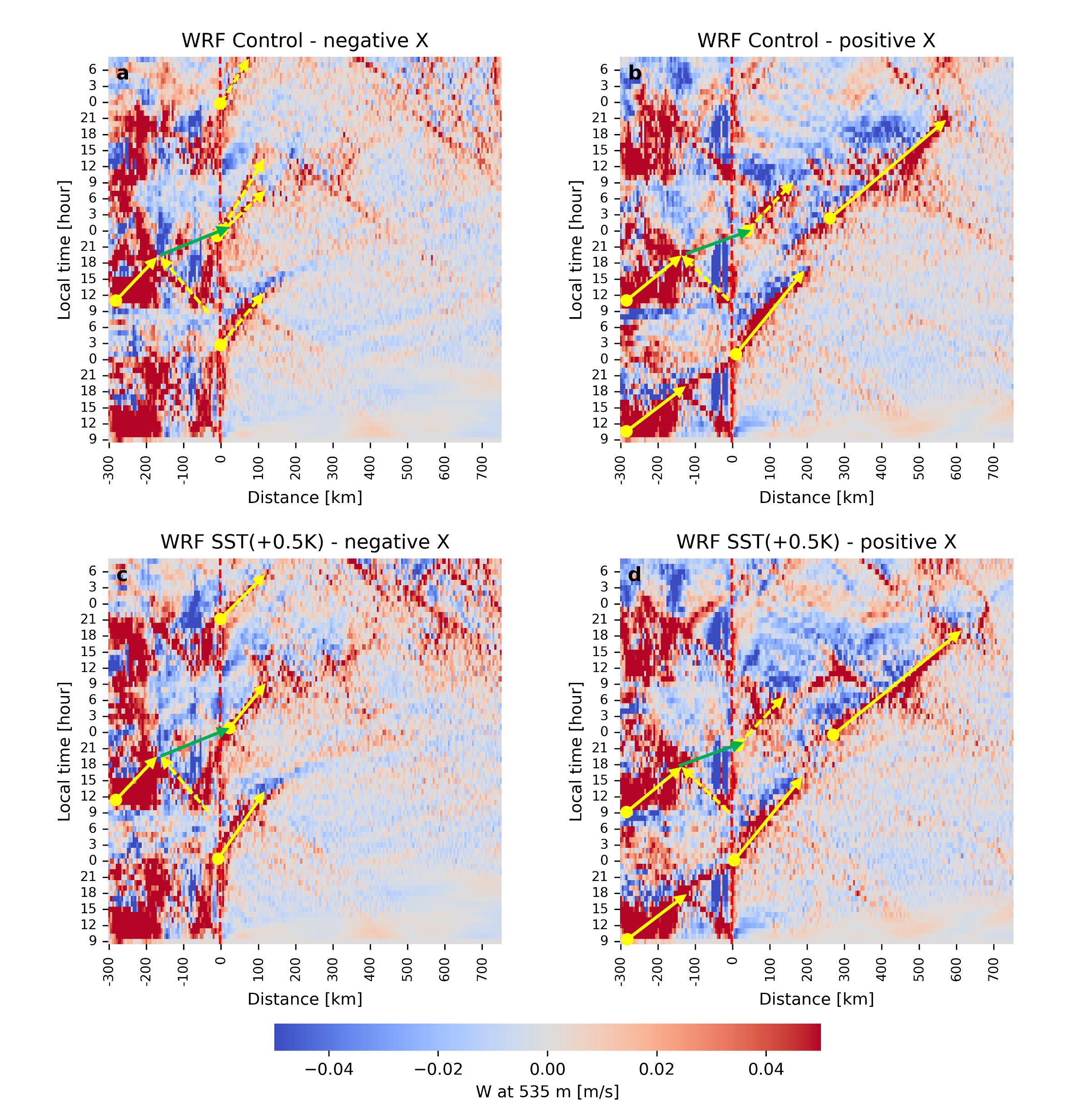}\\
 \caption{Similar to Figure \ref{fig:Hov.W.10km}, but at an altitude of 535 m (within the boundary layer). Solid and dashed yellow arrows represent relatively strong and weak convective updrafts, respectively, with yellow arrows featuring circular starting points indicating the three modes of convective offshore propagation. Green arrows represent the "jump" of convective signals between the first and second modes.}
 \label{fig:Hov.W.550m}
\end{figure}

Figure \ref{fig:Hov.W.550m} illustrates the vertical motion at an altitude of 535 meters, predominantly situated within the atmospheric boundary layer. Unlike in the free atmosphere, convective updrafts at this level are consistently paired with propagating convection. The convective updrafts associated with the second mode appear as two weak branches in the control run but merge into a single, stronger branch in the SST (+0.5 K) sensitivity experiment in the negative X channel on the second day. Pronounced downdrafts closely follow these updrafts, commonly persisting over the local ocean for more than 6 hours and over the local land for more than 15 hours. Over land, relatively weaker downward velocities at this height occur where two daytime opposing updrafts converge, whereas stronger and more persistent downdrafts are found closer to the coastal gap region between the two modes of convection. In particular, within the gap region, strong downdrafts persist for 10 hours, from 12 PM to 10 PM LST, within about 80 km onshore (Figure \ref{fig:Hov.W.550m}, right).

Furthermore, Figure \ref{fig:Hov.W.10km} indicates that during the second diurnal cycle of convection, particularly in the negative X channel, higher SSTs are associated with stronger updrafts that originate from the warmer ocean surface. In the control run, the updrafts tend to be weaker and shallower (Y between 100 to 250 km, day+2, from 9 AM to 6 PM), resulting in a bifurcation of the updraft structure. This suggests that SST is a key environmental factor influencing the renewal of the second mode by modulating the strength of convective updrafts. Apart from the differences in propagation speed, the gravity-wave signals in the free troposphere (Figure \ref{fig:Hov.W.10km}a–d), characterized by the alternating light red and light blue bands, exhibit relatively weak sensitivity to SST. In contrast, the boundary-layer convective updrafts (Figure \ref{fig:Hov.W.550m}, left), represented by the narrow dark red bands, show a more pronounced response to the SST+0.5 K perturbation. Therefore, the following results focus on the contribution of lower-tropospheric thermodynamics to convective offshore propagation.

\begin{figure}[t]
 \noindent\includegraphics[width=29pc,angle=0]{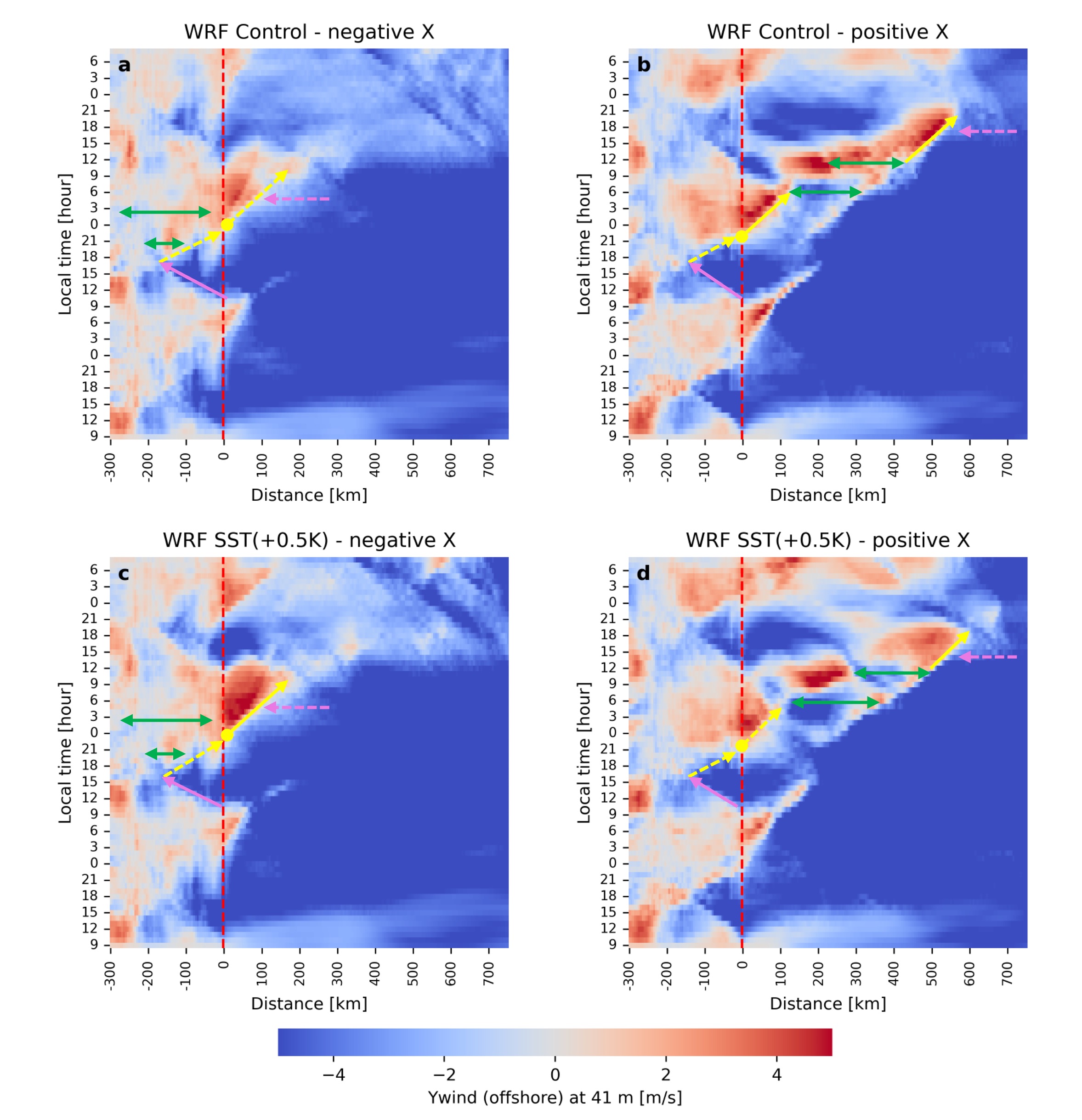}\\
 \caption{Similar to Figure \ref{fig:Hov.W.10km}, but showing offshore wind speed (m s$^{-1}$), which is the projections of U and V wind components along the y-axis, at an altitude of 41 m (near the surface). Solid and dashed yellow arrows represent relatively strong and weak offshore winds, respectively, corresponding to the hybrid land breeze gust front. Solid and dashed pink arrows indicate the onshore sea breeze front and background winds, respectively. Green bidirectional arrows depict localized cold pools that generate rapidly spreading winds in opposite directions.}
 \label{fig:Hov.Ywind.surface}
\end{figure}

Figure \ref{fig:Hov.Ywind.surface} illustrates the projection of the near-surface horizontal wind onto the y-axis, with positive (negative) values indicating offshore (onshore) winds. Within 150 km of the coastline, there is a noticeable diurnal variation. Supported by the background flow, the sea breeze gust front advances onshore at approximately -3 m s$^{-1}$, with embedded wind speeds surpassing -4.5 m s$^{-1}$. Around 4 PM, a marked transition in wind direction is observed, changing abruptly from onshore to offshore. This shift coincides with the convergence of the first mode and the sea breeze, leading to localized convection and the development of divergent flows characterized by opposing wind directions. This suggests that cold pools over land may play a role in this sharp transition in wind direction. However, given that this event occurs close to sunset, radiative cooling may also influence this modulation. 

The transition of wind speed reaches the coastline between 9 PM and 12 AM, ultimately developing into a land breeze that moves offshore. The movement, speed, and trajectory of the gust front align with the locations of precipitation. Notably, within the gust front of the land breeze, offshore wind speeds significantly increase upon crossing the coastline. This phenomenon is especially pronounced in the SST (+0.5 K) sensitivity experiment, where offshore wind speeds within the land breeze over the ocean typically reach up to 4.5 m s$^{-1}$, with peak values occurring along the gust front, coinciding with areas of convective precipitation. In contrast, offshore wind speeds in the control run remain around 2 m s$^{-1}$ before approaching the coastline and over the ocean. 

The diurnal cycles of vertical velocity and boundary-layer wind speed (Figures \ref{fig:Hov.W.10km}-\ref{fig:Hov.Ywind.surface})are strongly correlated with the convective offshore propagation signal (Figure \ref{fig:Hov.PR-GPM_C1_SST}). Over land, the diurnal variations of wind and propagation are relatively stable and linear, whereas over the ocean, they become more complex, exhibiting long-distance and discontinuous propagation, particularly in the positive X channel (Figure \ref{fig:Hov.Ywind.surface}). Meanwhile, the cold pools thought to be important are much smaller in scale than the northeastern coastline and require more detailed dynamical and thermodynamic analyses to determine whether the offshore propagation indeed originates from convection itself. Therefore, Hovmöller diagrams averaged along the X-axis alone are insufficient to fully test the cold-pool-based physical hypothesis or to explain the mechanisms underlying the oceanic propagation shown in Figure \ref{fig:Hov.Ywind.surface}b, highlighting the need for a more comprehensive examination of boundary-layer dynamics and moist thermodynamics.

\subsection{Moist boundary layer dynamics} 

\subsubsection{Afternoon to midnight: Convective jump and wind reversal}

\begin{figure}[t]
 \noindent\includegraphics[width=39pc,angle=0]{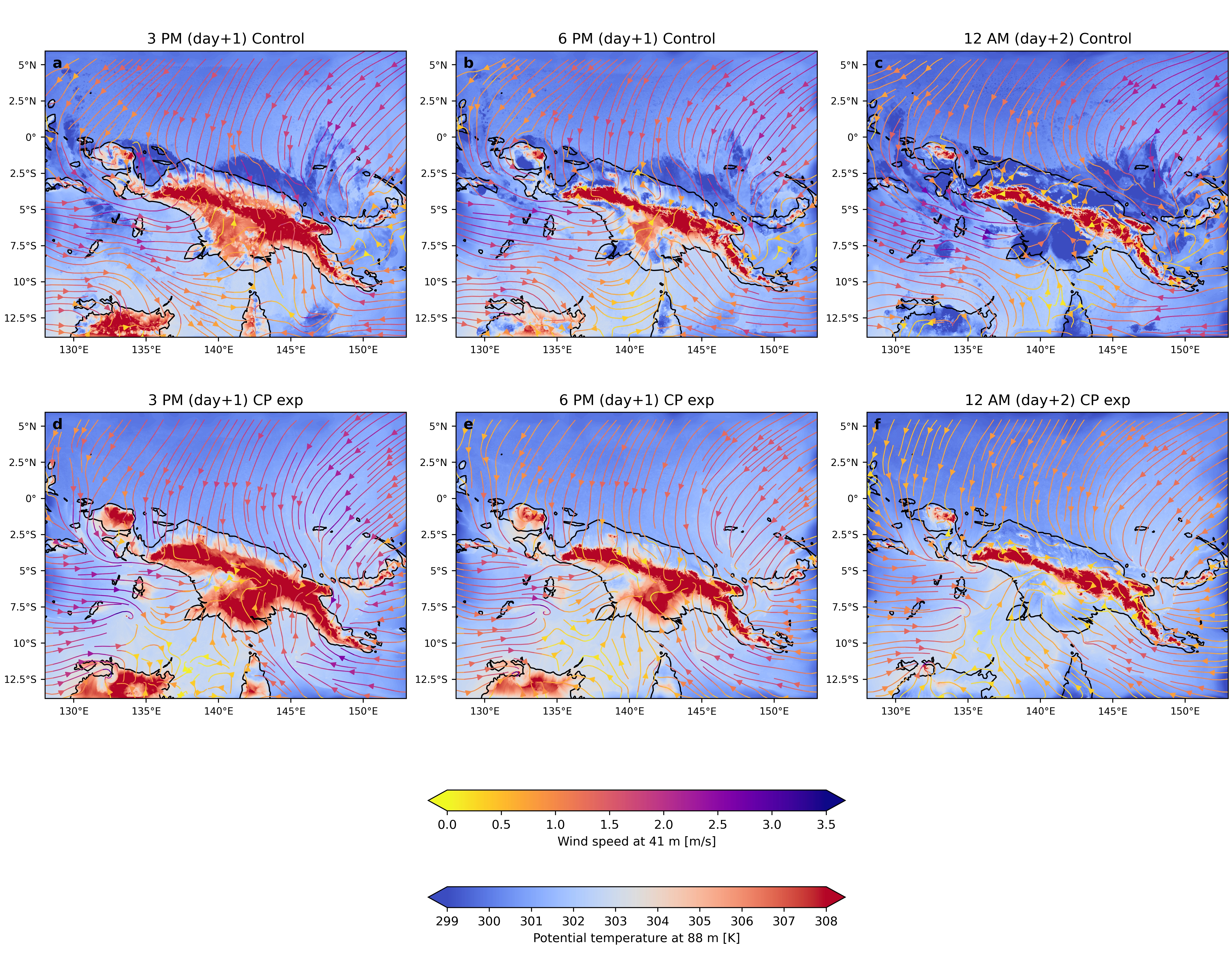}\\
 \caption{Snapshots of potential temperature [K] at 88 m altitude (colored). Wind streamlines represent horizontal wind at 41 m altitude, colored by wind speed magnitude. The left, middle, and right panels correspond to 3 PM (day+1), 6 PM (day+1), and 12 AM (day+2), respectively, in the control run (top) and the cold pool experiment (bottom).}
 \label{fig:snap.Tz2}
\end{figure}

In this section, we revisit the different stages within the second diurnal cycle to examine and substantiate the contributions of various types of density currents to propagating convection at each stage. We first focus on the period from 3 PM (day +1) to 12 AM LST (day +2), when first-mode convection develops over land, accompanied by the inland sea breeze in the afternoon, followed by a complete reversal of near-surface winds to a land breeze near the coast, along with second-mode convection initiated over the ocean.

Figure \ref{fig:snap.Tz2} illustrates snapshots of potential temperature and horizontal winds in the boundary layer from the afternoon to midnight for both the control run (a-c) and the cold pool experiment (d-f). These images highlight different types of density currents characterized by colder anomalies in the boundary layer. In comparison to the control run, the removal of evaporative cooling completely eliminates the offshore convection over the positive X channel, which persisted and intensified into the following day, along with the associated density currents. The sea-breeze edges are identifiable as sharp near-surface potential temperature gradients separating the relatively cooler marine air (blue shading) from the warmer land surface (red shading), with the cooler air advancing inland during the afternoon. Moreover, in the absence of cold pools, these edges become more clearly aligned with the coastline (whether straight or curved) between 3 PM and 6 PM, exhibiting a more coherent onshore propagation.

From 3 to 6 PM, sea breezes persist until just before sunset (Figures \ref{fig:clim.Hov.coast} and \ref{fig:Hov.Ywind.surface}), propagating across the island and triggering new convection with small cold pools at their leading edge (Figure \ref{fig:snap.Tz2}a-b), forming a distinct thermal boundary that runs roughly parallel to the complex shape of the coastline (Figure \ref{fig:snap.Tz2}d-e). The arrival of the sea breeze brings cooler air from the ocean, which makes it hard for the first mode of convection to continue its propagation toward the coastline, especially as land cools even further after sunset.

At 12 PM, the background wind over the northeastern region of the island closely resembles the spatial pattern of the February climatological cross-equatorial monsoonal flow (Figure \ref{fig:clim.Feb.Aug.mtn}b), which has difficulty penetrating the island's interior. From 3 PM to 12 AM, the onshore background wind speed gradually diminishes until it slightly reverses as the observed land breezes in some areas. 

The spatial scales and structures of land-sea breezes differ markedly from those of cold pools created by evaporative cooling over land. The former exhibit a spatial scale that corresponds with that of the island and remain nearly parallel to the intricate shape of the coastline. In contrast, cold pools are smaller and have more compact and circular features. These cold pools can either align with slightly cooler signals at the edges of the sea breeze or continue to develop overnight, thereby greatly enhancing the ability of the hybrid land breeze to drive convection offshore (the definitions of the land/sea breezes in Section 2.3). From 6 PM to 12 AM, the gust fronts of cold pools reach or even extend beyond the coastline into the ocean, particularly along the southwestern shore (Figure~\ref{fig:snap.Tz2}c), enhancing the development of offshore winds at night (cf. Figures~\ref{fig:snap.Tz2}c and \ref{fig:snap.Tz2}f).

\subsubsection{Early morning to after sunrise: Sustained offshore propagation}

\begin{figure}
 \noindent\includegraphics[width=39pc,angle=0]{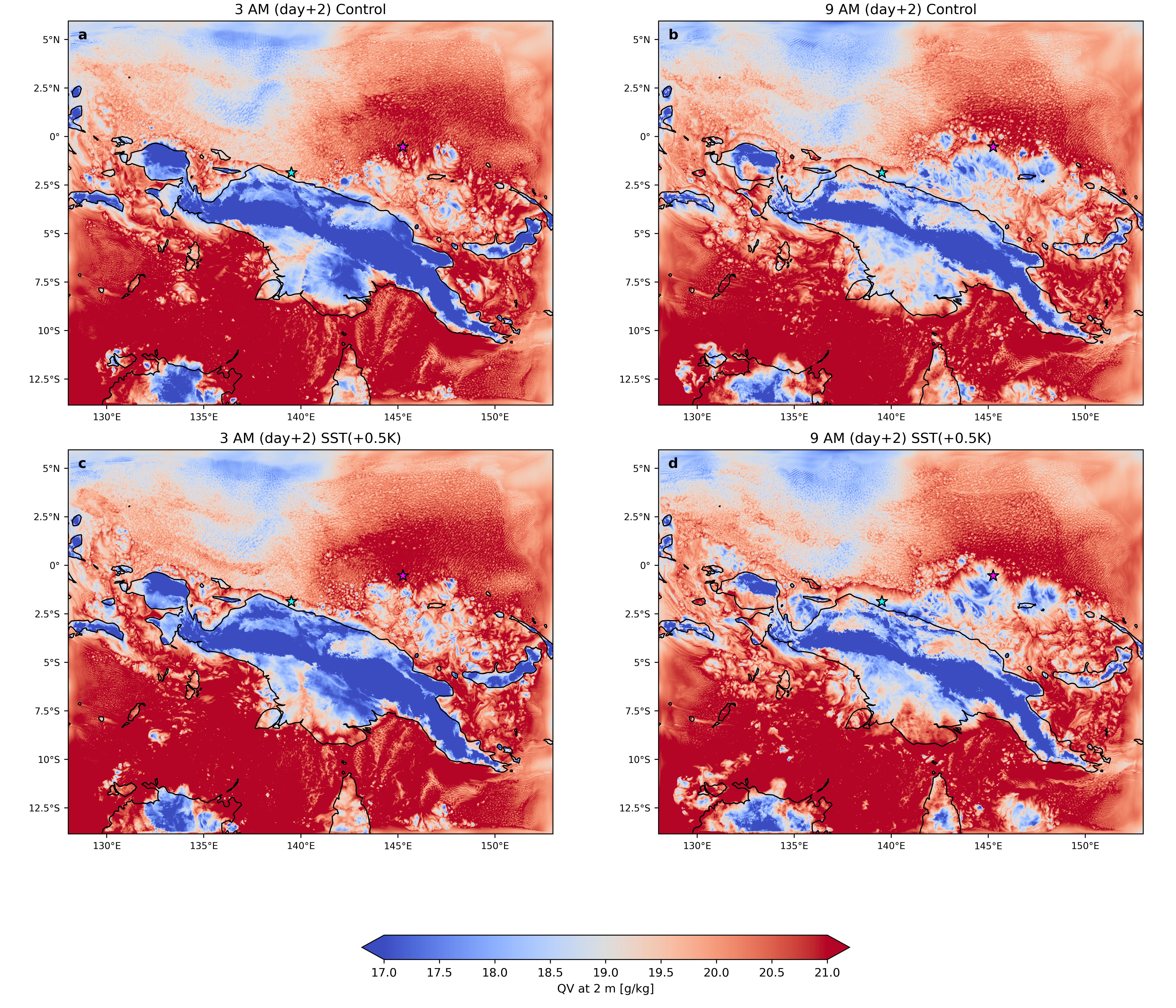}\\
 \caption{Snapshots of water vapor mixing ratio (colored) at 2 m altitude (near the surface) at 3 AM (day+2) (left) and 9 AM (day+2) (right) in the control run (top) and the SST (+0.5 K) experiment (bottom). The cyan star marks the land breeze (LB) point within the moist patches at 3 AM in the control run, coinciding with the land breeze gust front. The magenta star marks the cold pool (CP) point at 9 AM in the control run, coinciding with the moist patches at cold pool gust front and representing the farthest extent of offshore convective precipitation in the positive X channel.}
 \label{fig:snap.Q2}
\end{figure}

\begin{figure}
 \noindent\includegraphics[width=39pc,angle=0]{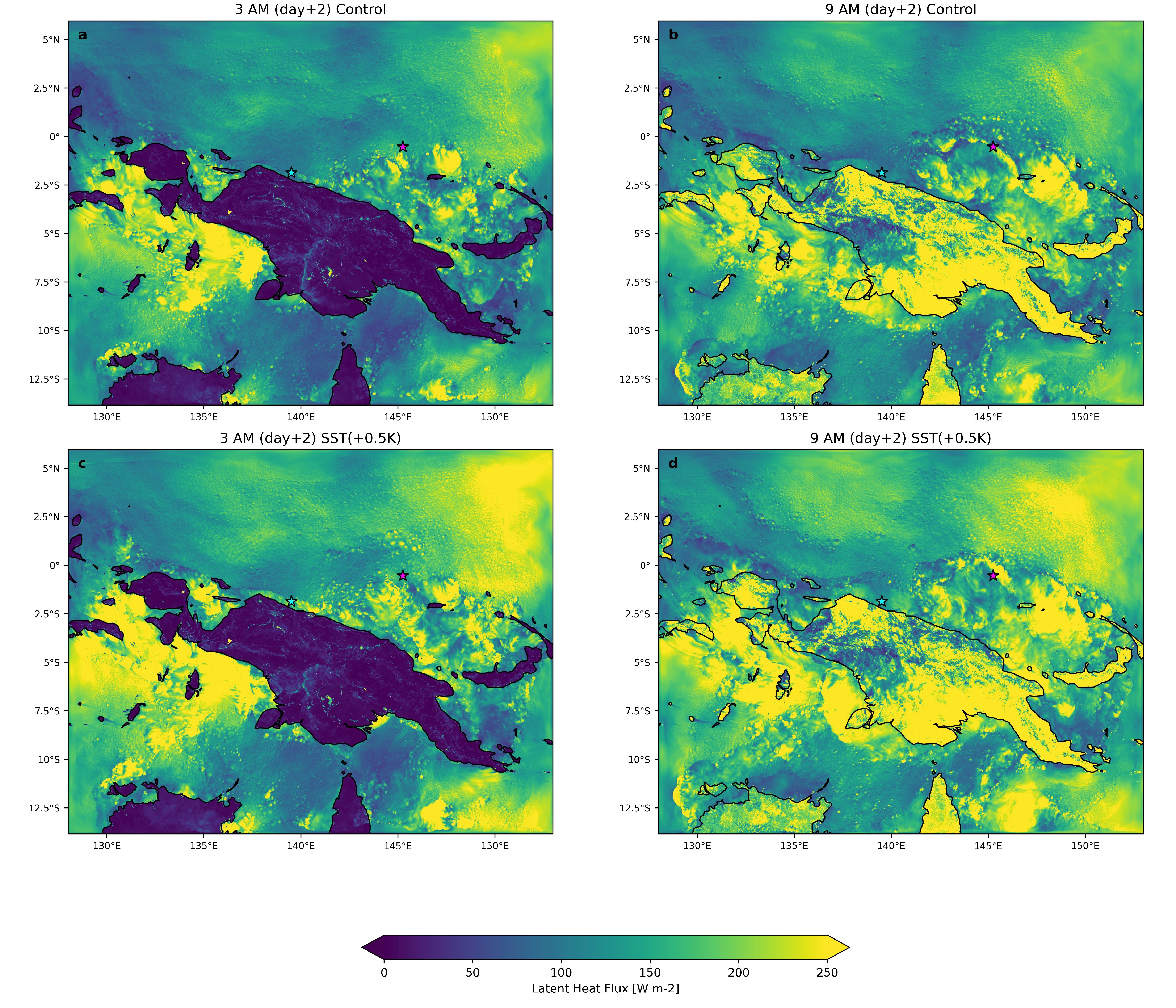} \\
 \caption{Similar to Figure \ref{fig:snap.Q2}, but showing latent heat fluxes (colored).}
 \label{fig:snap.LH}
\end{figure}

Between 3 AM and 9 AM (day +2), the second-mode convection remains organized and continues to propagate offshore over the ocean, even showing no immediate decay after sunrise. To better appreciate how convection propagates offshore, we consider snapshots of the near-surface water vapor mixing ratio (Figure \ref{fig:snap.Q2}) and surface latent heat fluxes (Figure \ref{fig:snap.LH}) at 3 AM (left panels) and 9 AM (right panels) of the second day. Throughout this period, the most distinct moist patches align well with the gust fronts of density currents and the secondary mode of convection (see Figures S5 and S6 in Supporting Information), marked by high water vapor content along the boundaries of density currents (Figure \ref{fig:snap.Q2}) and high latent heat fluxes within the density currents (Figure \ref{fig:snap.LH}). These moist patches predominantly form over ocean rather than over the land surface. In the SST experiment (+0.5 K) (c,d), the boundaries of moist patches are more pronounced compared to the control run (a,b).

In the control run (Figures \ref{fig:snap.Q2}a,b), numerous small circular cold pools generated by the previous day's offshore convection persist over the ocean between 3 AM and 9 AM (day +2), further organizing and enhancing convection. By 9 AM, these cold pools coalesce into a larger structure, within which conditions are drier than the surrounding areas and, thus, are also characterized by higher latent heat fluxes (Figure \ref{fig:snap.LH}b). A narrow but pronounced band of \textit{moist patches} emerges along the edges of the cold pools, representing a typical feature closely linked to the evolution of density currents and moist thermodynamics. We mark the farthest offshore point of this structure with a magenta star, referred to as the \textit{CP point}, which serves as a reference for subsequent cross-sectional analyses of the types, structures, evolution, moist thermodynamic effects, and their roles in initiating and propagating new convection.

Interestingly, moist patches are observed not only along the edges of cold pools generated by precipitation but also along the \textit{land breeze} front, extending parallel to the coastline at 3 AM (Figure \ref{fig:snap.Q2}a). We designate one such location with a cyan star, referred to as the \textit{LB point}. On the eastern side of the island, these moist patches along the land breeze front continue to propagate forward even after sunrise (Figure \ref{fig:snap.Q2}). When comparing the relative positions of the moist patches to the two star symbols, higher SSTs are found to be associated with stronger precipitation rates and convective updrafts (Figures \ref{fig:Hov.PR-GPM_C1_SST} and \ref{fig:Hov.W.550m}) as well as with faster and broader propagation of moist patches (Figure \ref{fig:snap.Q2}).

\subsubsection{From nighttime onward: Structure and evolution of density currents}

\begin{figure}
 \noindent\includegraphics[width=29pc,angle=0]{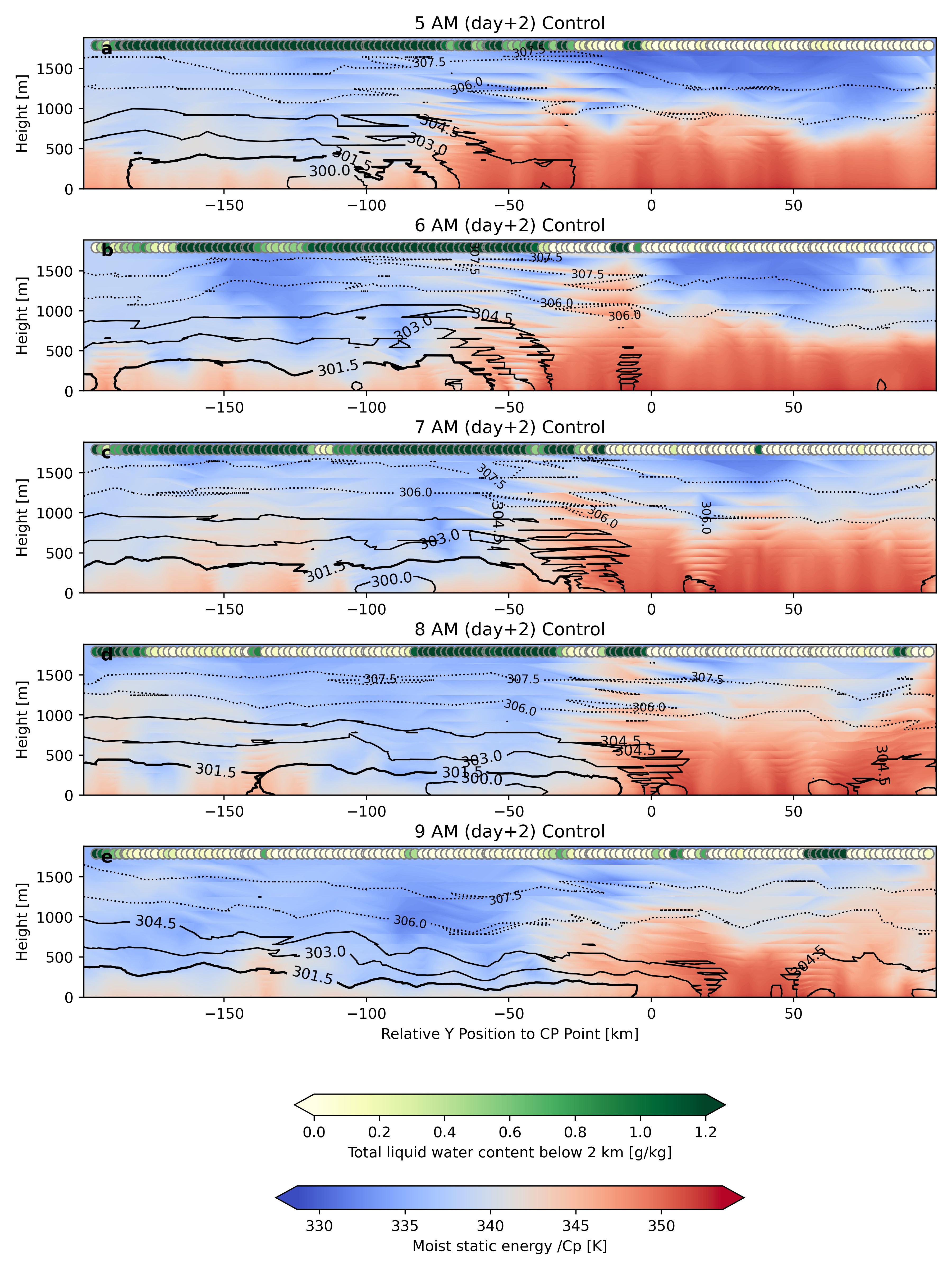}\\
 \caption{Cross-sections parallel to the Y-axis in Figure 1b (perpendicular to the coast) below 2 km, taken hourly from 5 AM to 9 AM (day+2). The horizontal axis is centered at the CP point (marked by the pentagram in the right panel of Figure 10), and values indicate the relative y-coordinate distance [km] from the CP point. Contours represent density potential temperature [K], where solid lines indicate the interior of boundary layer density currents (300–304.5 K), and dashed lines represent the exterior. The 301.5 K isotherm is thickened for clarity. The red and blue shading represents moist static energy divided by Cp (specific heat at constant pressure), with units in K. The scatter plot at the top of each panel shows column-integrated total liquid water content [g/kg] below 2 km, including both cloud water mixing ratio and rain water mixing ratio.}
 \label{fig:cross.CPP}
\end{figure}

The horizontal propagation of gust fronts depends on the density current’s thickness, the height of its momentum source, the origin of cooling, and the underlying moisture and flux conditions. These factors are best assessed using cross-sectional analyses. This section examines how the hybrid land breeze (see Section 1) in the negative X channel and the cold pools in the positive X channel, both key contributors, shape moist boundary-layer dynamics through their physical structures.

Figure \ref{fig:cross.CPP} illustrates cross-sections parallel to the y-axis within the coastline-aligned coordinate system at altitudes below 2 km. These sections are taken at the CP point, which corresponds to the location of the magenta star in the control run. The cross-section is centered at this point, with an averaging width of 1 km on either side along the coastline direction.

We use three key $\theta_\rho$ isopleths to map cold pool structure in Figure \ref{fig:cross.CPP} (see Section \ref{definitions} for details). The 300 K contour highlights the cold pool core, an area of particularly high density, formed by evaporative cooling beneath strong precipitating downdrafts. The 301.5 K isopleth defines the cold-pool boundary during direct pool–pool interactions ($Y < 0$ km before 08 AM LST), enclosing a layer approximately 450 m deep. In contrast, the 304.5 K contour traces the full density current front—and, for $Y > 0$ km before 08 AM LST, the mature cold pool interacting with the monsoonal flow—spanning roughly 1 km in depth.

Between 7 AM and 9 AM (day +2), the presence of a precipitation-driven downdraft, characterized by a negative MSE anomaly and a downward-branching structure that serves as the source of cold pools, can be inferred roughly between -100 km and -50 km. The spatial extent and the magnitude of total liquid water content associated with the old convective cell gradually diminish, while the cold pool drives continue to expand before weakening. At 8 AM, this fully developed cold pool reaches its maximum size (-140 to -10 km) to a radius that we estimate at approximately 65 km, and its left boundary appears to intersect with another cold pool. Below 750 m, this interaction leads to the accumulation of higher MSE within the 301.5 K contour (Y between –75 and 20 km from 7 to 9 AM; Y between –160 and –120 km at 8 AM) and is accompanied by new convective cells generated along the gust front, characterized by relatively high total liquid water content (Y between –20 and 20 km from 7 to 9 AM; Y at around –140 km from 8 to 9 AM). Elevated MSE is also present within the cold pools, peaking at the gust front and extending up to about 450 m above the upper boundary of the cold pool. Furthermore, the highest MSE within the cold pool is concentrated closer to the ocean surface.

A pronounced mid‐level monsoonal head (with surface signal around $(1^\circ\mathrm{N},\,146^\circ\mathrm{E})$ in Figure \ref{fig:snap.Q2}, with dynamics in Movie S2 in Supporting Information, illustrated in Section \ref{sec:monsoon}; flow in Figure \ref{fig:snap.Tz2}) lies immediately east of the cold‐pool front, exhibiting a distinct thermodynamic profile before interaction at 05 AM (Figure \ref{fig:cross.CPP}). In this study, we interpret the cross-equatorial monsoon intrusion as a large scale density current centered near 1–2 km, whose leading head interacts with the shallower density currents associated with cold pools and the land breeze. The 306.0 K and 307.5 K $\theta_\rho$ isopleths curve downward between –25 km and +100 km, and below the 306 K contour MSE increases sharply. This suggests that the air between 800 and 2000 m is markedly denser, colder, and drier than the surrounding environment, which can suppress convective development in the warm, moist boundary layer below 800 m. This layer corresponds to the dynamic core of the monsoonal flow, typically centered near 1500 m in the lower troposphere—well above the near-surface core of small-scale density currents.

Between 5 AM and 9 AM, differences in MSE and $\theta_\rho$ between the monsoonal air and its surrounding environment become less distinct. This indicates that cold pool gust fronts lift and mix warm, moist boundary layer air with the drier monsoonal air with the help of newly generated convection, thereby blurring the boundary between the monsoon layer and the ambient environment and weakening the thermodynamic contrast that initially defines the monsoonal head. This uplift generates new convective cells along and beneath the leading edge of the monsoon layer. By entraining higher-MSE air aloft, the gust fronts progressively warm and moisten the monsoonal core—enhancing its instability and favoring further convection below. Large, mature cold pools generated by the second mode of organized convective activity coexist with developing smaller cold pools in advance (Y = between –10 to 100 km). Finally, precipitation in the positive X channel and the density current (marked by the 304.5 K $\theta_\rho$ isopleth) move offshore at the approximate speed of 4.6 m s$^{-1}$. While the propagation and interaction of density currents are the primary drivers of precipitation, the local persistence of old convective cells is noticeably shorter than that of the density currents themselves. Nevertheless, density currents continuously generate new convective cells at their gust front, leading to the discontinuous and intermittent nature of convective precipitation.

\begin{figure}
 \noindent\includegraphics[width=39pc,angle=0]{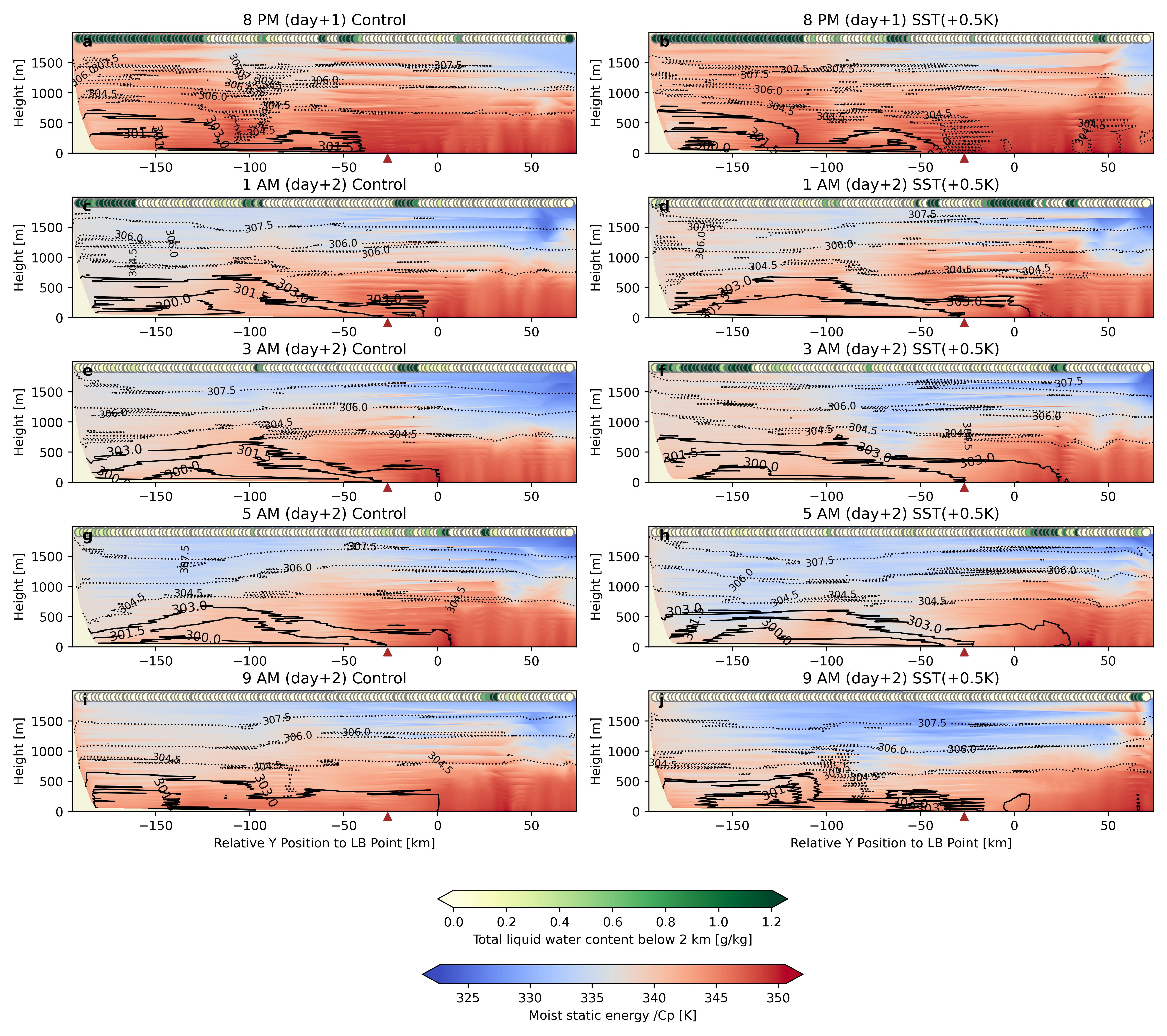}\\
 \caption{Similar to Figure \ref{fig:cross.CPP}, but comparing the control run (left) and the SST (+0.5 K) experiment (right) at 8 PM (day+1), 1 AM (day+2), 3 AM (day+2), 5 AM (day+2), and 9 AM (day+2). The range of solid-line contours is adjusted to 300–303 K, but they still represent the interior of density currents in terms of density potential temperature [K]. The brown triangles indicate the location of the coastline. Beige areas indicate missing data caused by the mountain.}
 \label{fig:cross.LBP}
\end{figure}

Figure \ref{fig:cross.LBP} shows cross-sections at the LB point for both the control run and the SST (+0.5K) experiment, from 8 PM (day +1) to 9 AM (day +2). After sunset, radiative cooling over land initiates a land breeze density current (see Section \ref{definitions} for details), identified here using the 303 K $\theta_\rho$ isopleth, which moves offshore until it dissipates after sunrise the following day. The land breeze is shallower than cold pools (Figure \ref{fig:cross.CPP}), exhibiting a thickness of 250-400 m over the ocean, and it also has lower MSE anomalies. Consistent with our hypothesis, the hybrid land breeze appears to be thicker over land, particularly in areas experiencing precipitation, reaching thicknesses of up to 600 m.

By 8 PM LST on day +1, the first convective mode produces a broad cold pool along the main ridge (Y = between –200 to –125 km; Figure \ref{fig:cross.LBP}, referring to Figure \ref{fig:snap.Tz2}b-c), while afternoon sea‐breeze cells trapped by coastal hills in the negative-X channel remain between Y = –80 and –40 km, yielding smaller cold pools. At Y $\sim-$100 km, these flows interact by 8 PM, and from 8 PM to 1 AM LST precipitation-drive downdrafts inject low MSE air from aloft. Between 1 and 5 AM LST, the resulting density currents stabilize: the 301.5 K $\theta_\rho$ contour stays fixed at Y $\sim-$100 km, marking both the organized nocturnal cold pool and intermittent foothill pools ($\theta_\rho$ $\sim$ 300 K), whereas the 303 K contour alone tracks the advancing land-breeze front over the ocean.

The land breeze front advances at a nearly uniform speed of approximately 1.4 m s$^{-1}$ before sunrise, accompanied by the high MSE values along its leading edge and new convective cells. From 8 PM to 5 AM, the location of convection at -100 km remains relatively stable, with new convective cells forming only offshore in both the control run and the SST (+0.5K) experiment (Figure \ref{fig:cross.LBP}). This process creates moist patches that enhance the buoyancy of surrounding air parcels, subsequently regenerating offshore propagation as the second mode.

Figure \ref{fig:cross.LBP} also reveals mid-level monsoonal air adjacent to the density-current fronts. In the positive X channel (Figure \ref{fig:cross.CPP}), a layer of relatively cold and dry monsoonal air is located on the offshore side of the density current and extends eastward along the coastline. In the negative X channel (Figure \ref{fig:cross.LBP}), this layer reaches closer to the coast due to the recurvature of the cross-equatorial flow toward the island. Above 1 km, the presence of this dry layer initially suppresses low-level convection (Figure \ref{fig:cross.LBP}a–b). As the land-breeze gust front advances offshore, it lifts and warms the monsoonal air, mixing it with higher-MSE air aloft and triggering shallow convective cells that persist through sunrise (Figure \ref{fig:cross.LBP}c-h). By 9 AM  LST, increasing latent heat fluxes \cite{grant2016, tang2024} and the return of the daytime sea breeze dissipate the outflow buoyancy, causing the front to retreat onshore (cf. Figure \ref{fig:cross.LBP}g–i, h–j).

Compared to the control run (Figure \ref{fig:cross.LBP}, left), the SST (+0.5K) experiment (Figure \ref{fig:cross.LBP}, right) shows a similar pattern. Despite a 0.5 K increase in SST, the boundary defined by the 301.5 K $\theta_\rho$ isopleth remains fixed at the coastline, mirroring the control run. Compared to the control, the gust front in this sensitivity experiment leads by approximately 25 km and is associated with a larger area of liquid water content. This difference could be attributed to an earlier initiation of the land breeze due to the increase in the land-sea contrast. In addition, higher SSTs might lead to greater evaporative fluxes and precipitation rates, which in turn would intensify downdrafts and the following cold pools. This would be reflected in stronger offshore winds within the gust front of the hybrid land breeze with higher SSTs (between 0 and 200 km offshore)(Figure \ref{fig:Hov.Ywind.surface}, left). However, due to higher latent heat fluxes (Figure \ref{fig:snap.LH}), the land breeze dissipates more readily in the SST sensitivity experiment.

\section{Discussion}
\label{sec:discussion}

Nocturnal radiative cooling generates land breezes, and evaporative cooling from precipitation leads to the formation of cold pools, both of which are key sources of density currents in the boundary layer and are widely recognized for their role in organizing and propagating convection. The general phase-locked mechanisms over New Guinea found by our study may operate over other tropical islands and coastal regions, although successful regeneration of an over-ocean mode requires sufficiently warm sea surface temperatures and abundant atmospheric moisture. The convective systems examined in this study are predominantly organized as propagating lines of convective cells, yet they differ from typical squall lines that develop under strong low-level wind conditions. This distinction arises because the mountain ridge and the northeastern coastline of New Guinea, which initiate and regenerate the first and second convective modes, respectively, are nearly linear in shape, providing the basis for the two quantitative coordinate axes defined in this study.

The direction and speed of diurnal convective propagation are strongly controlled by inland and offshore winds in the boundary layer (Figure~\ref{fig:Hov.Ywind.surface}), which distinguish both climatological-scale events extending over approximately 200~km (Figure~\ref{fig:Hov.Ywind.surface}a, consistent with observations in Figure \ref{fig:clim.Hov.coast}) and exceeding 600~km offshore (Figure~\ref{fig:Hov.Ywind.surface}b; schematically illustrated in part of Figure~\ref{fig:moistBL.schematic}). During the day, convection develops over elevated terrain and subsequently propagates fast toward the coast. The associated cold outflows or gust fronts (Figures \ref{fig:Hov.W.550m}--\ref{fig:snap.Tz2}) accelerate the transition from the prolonged sea breeze ($-3.7~\mathrm{m~s^{-1}}$) to an earlier onset of the observed land breeze ($2~\mathrm{m~s^{-1}}$) before 6 PM LST. At night, the combined effects of radiative cooling and residual cold pools initiate a hybrid land breeze whose leading front with embedded moist patches (Figures \ref{fig:snap.Q2}--\ref{fig:snap.LH}), advancing at approximately $2.5$ to $4.5~\mathrm{m~s^{-1}}$, triggers new offshore convection—the second mode. Over warmer ocean surfaces (Figure~\ref{fig:Hov.Ywind.surface}c compared with Figure~\ref{fig:Hov.Ywind.surface}a), this second-mode convection produces more intense cold pools that merge with the body of the land breeze, increasing its internal flow speed to $4.5$ to $7.2~\mathrm{m~s^{-1}}$ and driving both the land breeze front and newly generated convection offshore at nearly $4~\mathrm{m~s^{-1}}$ (supported by Figure~\ref{fig:cross.LBP}).

When the convection and cold pools initiated by the land breeze are sufficiently strong and extensive, the resulting positive feedback described above can \textit{overcome the regular diurnal cycle}, continuously generating new convection and cold pools that sustain offshore propagation and maintain moist patches. As a result, the convection maintains a long-lived offshore propagation dominated by moist-boundary-layer density currents (cf. Figures~\ref{fig:Hov.Ywind.surface}b and a; cf. Figures \ref{fig:snap.Tz2} a-c and d-f; supported by Figures~\ref{fig:snap.Q2}--\ref{fig:cross.CPP}). In this case, the convection initiated on the first day does not decay with the land breeze but persists over the ocean (Figure~\ref{fig:Hov.PR-GPM_C1_SST}d). A subsequent land breeze generated on the second day further enhances convection and cold pools, merging with the preexisting offshore system and amplifying the positive feedback. Between the two MCSs and in their surrounding region, the offshore winds exhibit gust-front-like structures propagating in opposite directions, oscillating with amplitudes exceeding $\pm4.5~\mathrm{m~s^{-1}}$ (Figure~\ref{fig:Hov.Ywind.surface}b). This cross-day coupling, reinforced by cold pools, ultimately allows the organized convective system to propagate more than $600~\mathrm{km}$ offshore.

\subsection{Density currents vs. gravity waves}

Gravity waves and density currents have long been acknowledged as two significant yet often challenging processes contributing to convective offshore propagation \cite{houze1981monsoonOP, mapes2003GWop}. To differentiate their relative contributions, we examine differences in propagation speeds (whether they are faster or slower than 10 m s$^{-1}$) and signatures at various altitudes: gravity waves can be observed at the upper troposphere (Figure \ref{fig:Hov.W.10km}), whereas density currents are mostly confined in the boundary layer (Figure \ref{fig:Hov.W.550m}).

Gravity waves generated by diurnal and offshore-propagating convection travel at speeds markedly faster than the relatively slow offshore movement of the convective system itself (cf. Figures \ref{fig:Hov.W.10km} and \ref{fig:Hov.PR-GPM_C1_SST}). Furthermore, the \textit{sensitivity} of offshore-propagating convection to SST does not show corresponding variations in gravity wave signals (cf. Figure \ref{fig:Hov.W.10km} a-c, b-d). As a result, for these offshore-propagating MCSs, gravity waves are unlikely to be the primary mechanism driving their offshore movement, even in the simulated case where the system propagates over distances exceeding 600 km.

However, due to the complexity of overlapping gravity wave signals depicted in Figure \ref{fig:Hov.W.10km}, it remains challenging to definitively ascertain whether these waves originate from diurnal convection over land, diurnal radiative cooling and heating, or a combination of both. Even waves generated by the same source can produce different gravity wave modes with varying vertical wavelengths and propagation speeds. In our results, two prominent gravity waves propagate near 10 km altitude at approximately 18 and 9 m s$^{-1}$, respectively. They are excited by the first and second modes of convection and begin around 12 PM and 12 PM (day+1), respectively (Figure~\ref{fig:Hov.W.10km}), and there may even exist interactions that span across an entire diurnal cycle. Previous studies have hypothesized that upward motion and adiabatic cooling associated with gravity waves emanating from daytime land convection may reach far offshore by nighttime and trigger convection, thereby facilitating diurnal offshore propagation of precipitation \cite{love2011, hassim2016diurnal, yokoi2017, ruppert2020convGW}. However, such an effect is not evident in our simulations.

Relative to the free atmosphere, vertical motions within the boundary layer (Figure~\ref{fig:Hov.W.550m}), comprising organized convective updrafts and downdrafts, exhibit a coherent structure dynamically coupled with propagating convection. Density currents, including land breezes and cold pools, are also \textit{diabatically} driven motions but can also propagate \textit{horizontally} above the surface due to \textit{negative buoyancy}, continually generating new convective cells along their gust fronts (Figures~\ref{fig:cross.CPP}–\ref{fig:cross.LBP}). This provides theoretical support for the role of near-surface horizontal winds synchronized with gust fronts (Figure~\ref{fig:Hov.Ywind.surface}) in driving convective propagation and sustaining the long-lived nature of propagating convective systems (Figure~\ref{fig:Hov.PR-GPM_C1_SST}). The first mode of diurnal convection initiates over the central part of the island in the early afternoon (Figure~\ref{fig:clim.Feb.Aug.mtn}, left), producing cold pools that spread outward and move toward the coast. Subsequent nighttime radiative cooling over land generates land breezes that propagate seaward from the coast. Together, these two types of density currents can serve as basic diurnal drivers of the offshore propagation.

In the tropics, density currents operate not only through dry dynamics, such as mechanical lifting. Although environmental low-level wind shear is generally weaker than in the midlatitudes, the tropical oceans and the Maritime Continent feature very moist and warm boundary layers with strong conditional instability. This is reflected in our results , where the second mode shows sensitivity to SST perturbations(e.g., Figure~\ref{fig:Hov.Ywind.surface}) and depends on moist patches at the edges of density currents (e.g., Figure~\ref{fig:snap.Q2}). Therefore, tropical density currents, situated at more favorable altitudes within the moist boundary layer, are particularly effective at tapping into its energy and working in concert with \textit{moist thermodynamics} to sustain convective offshore propagation.

Using propagation speed alone to distinguish between gravity waves and density currents, however, may overlook important factors such as topographic effects, rapid jump process around 8 m s$^{-1}$ by sea breeze–cold pool–land breeze interference, the interaction between different types of density currents, and the influence of background winds. Notably, steeply sloped terrain can significantly accelerate the descending gust front of density currents, and with the assistance of nighttime katabatic winds, can impart an initial offshore propagation speed ranging from 8 to 11 m s$^{-1}$ (Figure \ref{fig:clim.Feb.Aug.mtn}a).

\subsection{Land breezes vs. cold pools and their moist patches}
\label{sec: hybrid LB}

Both land breezes and cold pools behave as density currents, and their propagation speed can be estimated using the following  \cite{benjamin1968}:
\begin{equation}
U \approx C \sqrt{g h \left(\frac{ \theta_{\rho,e}-\theta_{\rho}}{\theta_{\rho,e}}\right)},
\end{equation}
where \( U \) is the density current velocity, \( h \) their height, and \( C \) is a dimensionless empirical coefficient set to 0.6 for the tropical ocean surface. Estimated propagation speeds for land breezes and cold pools in our study, based on their vertical structure and surrounding environment (Figures \ref{fig:cross.CPP}-\ref{fig:cross.LBP}), are summarized in Table \ref{tab:propagation-speeds}. These estimates are generally consistent with our simulated results (Figure \ref{fig:Hov.Ywind.surface}a), although minor discrepancies exist. For example, strengthened by background winds and warm sea surface temperatures, the land-breeze front can reach speeds exceeding 4.5~m~s$^{-1}$ over the ocean—intermediate between the typical propagation speeds of land breezes and convectively generated cold pools.

\begin{table}[htbp]
\centering
\caption{Estimated propagation speeds for land breeze and cold pools.}
\begin{tabular}{lcccc}
Density Current & $h$ (m) & $\theta_{\rho}$ (K)& $\theta_{\rho,e}$ (K) & $U$ (m\,s$^{-1}$) \\
Land breeze & 250 & 303  & 304.5 & $2.1$ \\
Cold pool & 1000 & 301.5 & 306 &  $7.2$ \\
\end{tabular}
\label{tab:propagation-speeds}
\end{table}

As the land breeze is treated here as a density current driven by radiative cooling over land, it propagates seaward from the coast. The over-ocean land breeze, which contributes to the propagation of the second mode, is notably faster than the $\sim$1.8 m s$^{-1}$ estimated from recent observations at Bengkulu Airport \cite{north2025}. In our framework, a similar weaker signal
 (Figure~\ref{fig:Hov.Ywind.surface}) instead represents the inland wind reversal when late-afternoon cold pools overcome the onshore sea breeze. 
Importantly, offshore winds over land on the inland side of the coastline differ fundamentally from the land breeze represented as a density current, highlighting that real islands cannot be idealized as spatially infinitesimal points and that land-breeze wind speeds are inherently nonuniform in space.
The simulated and theoretically estimated offshore wind speeds agree with shipborne observations from the R/V Mirai off the western coast of Sumatra, which recorded peaks exceeding 6 m s$^{-1}$ and a virtual potential temperature drop of 2–3 K \cite{peatman2023}. We suggest that various density currents, as well as the horizontal winds within their heads and bodies, evolve over time; hence, local wind speeds also change throughout the propagation process. This interpretation is consistent with the local observed gradual weakening of wind speeds to $\sim$2 m s$^{-1}$ after the peaks \cite{peatman2023}. These results highlight the need for future observations—simultaneous measurements at multiple locations, attention to gradual wind-speed changes  in addition to abrupt directional shifts, combined analyses with water vapor mixing ratio, and even Lagrangian storm-chasing approaches—to further test our simulations and the proposed convectively modified \textit{hybrid-land-breeze hypothesis}.

The hybrid land breezes in our simulations are not isolated entities, but interact with other dynamical features of deep convection. For example, the land breeze frequently triggers new convective cells that generate cold pools, which in turn supply cold air to the land-breeze current, aiding its maintenance and propagation. Over land, convection can last for several hours after sunset thanks to the interaction of the land breeze, cold pools, and mountain breezes with the possible help of radiative cooling (Figures \ref{fig:clim.Hov.coast} and \ref{fig:Hov.PR-GPM_C1_SST}). Consequently, large and deep cold pools over land can mix into the hybrid land breezes during nighttime (Figure \ref{fig:cross.LBP}).

To clarify the role of land breezes and cold pools in driving the offshore flow, we find that the observed signal results from a combination of both processes, consistent with our hypothesis. As shown in Figure \ref{fig:cross.LBP}, cold pools originating from inland convection merge with the land breeze, enhancing its strength and offshore extent. However, land breezes can form independent of cold pools. This can be seen in Figures \ref{fig:snap.Tz2}d–f, where a nighttime land breeze remains present even when cold pool effects are removed, as indicated by the reversal of flow direction compared to daytime. The spatial structure further supports this distinction: the land breeze appears as a narrow, coastline-parallel density current (Figure \ref{fig:snap.Q2}), in contrast to the arc-shaped, divergent structure of cold pools (Figures \ref{fig:snap.Tz2}b,c). While cold pools are not necessary for land breeze formation, they serve as an additional source for the density current, intensifying the offshore flow when present, as seen by comparing Figure \ref{fig:snap.Tz2}b,c and Figure \ref{fig:snap.Tz2}e,f. 

We further hypothesize that the incorporation of cold pools is a key factor enabling hybrid land breezes to propagate significantly beyond the traditionally expected 100 km. These cold pools effectively increase the overall thickness of the density currents and their gust wind speeds while also providing the thermodynamic and dynamic inertia characteristic of density currents (Figure \ref{fig:cross.LBP}). As a result, the hybrid land breezes exhibit extended duration and enhanced offshore propagation speed, allowing the diurnal cycle of convection to consistently reach offshore distances of 200–350 km (Figures \ref{fig:clim.Hov.coast} and \ref{fig:Hov.PR-GPM_C1_SST}).

Just like cold pools, our study shows that land breezes, too, can develop \textit{moist patches} at their gust front (Figure \ref{fig:snap.Q2}). These are well aligned with the coastline and coincide with the offshore propagation of the second mode of convection (Figure \ref{fig:cross.LBP}). They persist after sunrise the following day (Figures \ref{fig:snap.Q2} and \ref{fig:cross.LBP}) and are associated with elevated latent heat fluxes within the boundary layer (Figure \ref{fig:snap.LH}). Cold pools embedded in the land breeze can enhance the moist patches, as can higher SSTs. The large latent heat fluxes underneath the anomalous moisture (Figure \ref{fig:snap.LH}) support the notion that these moist patches arise from the propagation of the land breeze rather than being caused by moisture advection driven by the cross-equatorial monsoon. The latter primarily transports water vapor ahead of and beneath the dry, cold air masses, leading to large-scale moisture advection patterns (Figure \ref{fig:monsoon.schematic}b).

\subsection{Sea breezes and land breezes}

\label{concept:SCL interference}

In our study, sea breezes appear as moist, deep (550-1100 m), pressure-driven flows (see Figures S7 and S8 in Supporting Information) maintained by daytime heating and coastal topography. They advance slowly, with their fronts moving at an approximate speed of 3.7 m s$^{-1}$, even when onshore winds are taken into account (Figure \ref{fig:Hov.Ywind.surface}). In contrast, nocturnal land breezes are dry, shallow ($\sim$250 m), density currents, buoyantly accelerated by radiative cooling and often intensified by embedded cold pools, reaching speeds of 2–5~m~s$^{-1}$ at their gust front. Sea‐breeze dynamics appear akin to a near‐balanced flow regulated by thermal low‐pressure systems as well as buoyancy, whereas land breezes behave more like density currents driven primarily by cooling. Consequently, depth alone cannot predict propagation speed: deeper sea breezes may move more slowly than their shallower, buoyancy‐driven counterparts.

In our simulations, sea breezes typically develop around 10 AM and persist until about 6 PM (Figures \ref{fig:Hov.Ywind.surface} and \ref{fig:snap.Tz2}), consistent with previous studies \cite{miller2003SB, zhou2006NGtopoGW} and recent observations \cite{north2025}. This longevity results from intense and sustained radiative heating that maintains a thermal low over land. Once convection develops over land, the associated inflow arising from mass-conservation effects can further strengthen the sea breeze, consistent with the observed speed peak around 2 PM \cite{north2025}. After sunrise, while the sea-breeze signal becomes evident, the land breeze remains active offshore because its thermodynamic and dynamic inertia as a density current allows it to persist (Figure \ref{fig:cross.LBP}), gradually dissipating over the ocean by around 3 PM (Figure \ref{fig:Hov.Ywind.surface}). This persistence may obscure early-morning sea-breeze signals over land, although the two circulations often coexist during the morning hours.

Behind the sea-breeze head, colder air becomes trapped within the boundary layer over land during the afternoon and cools further after sunset, contributing to the 100 km shift between the first and second convective modes. Sea and land breezes also interact across both space and time, often through precipitation-driven cold pools that link their circulations. We refer to this coupled process as the \textbf{sea breeze–cold pool–land breeze interference} (Figures \ref{fig:clim.Hov.coast}, \ref{fig:Hov.PR-GPM_C1_SST}, and \ref{fig:cross.LBP}, as illustrated in Figure \ref{fig:moistBL.schematic}a-c, see Section \ref{definitions} for definitions), and we summarize it in three steps:
\begin{enumerate}
\item \textbf{Convergence and cold pool amplification.} The onshore sea breeze meets the first mode rain band just as the flow reverses into a land breeze, focusing moisture, triggering stronger cold-pool formation, and sustaining localized rainfall for hours.

\item \textbf{Boundary layer trapping and jump.} The sea breeze traps cold air from the coast inland before meeting propagating convection, stabilizing the boundary layer and halting the coastward advance of the first mode; hence the $\sim$100 km jump. The lack of precipitation there, combined with evaporation driven by low-level winds throughout the diurnal cycle (Figure \ref{fig:Hov.Ywind.surface}), may further dry the near-coastal area, serving as a local feedback.

\item \textbf{Merging.} Cold pools generated at the convergence zone quickly merge with the developing land breeze; however, due to insufficient surface heating, they are unable to regain buoyancy, thus suppressing the initiation of new convection ahead of the front until it reaches the coast.

\end{enumerate}

The interference described above amplifies the offshore propagation of the second convective mode. By merging with and strengthening the land‐breeze density current, originally driven by radiative cooling, cold pools enhance the gust front’s speed. Once this reinforced outflow clears the trapped cold‐air zone and taps into warm, moist ocean air, it regenerates new convective cells at the coastline, sustaining the second mode signal as it travels farther offshore.

The interference partly initiates when the sea‐breeze front begins its inland propagation, $\sim$12 PM in observations (Figure \ref{fig:clim.Hov.coast}) and $\sim$10 AM in the model (Figures \ref{fig:Hov.Ywind.surface} and \ref{fig:Hov.PR-GPM_C1_SST}). Roughly 2 hours later, the first mode of convection develops over land, marking the onset of full interference—consistent across both observation and simulation. The exact \textit{timing} of the subsequent steps, as well as the prominence and propagation distance of the first mode and the convective jump where the sea breeze traps cooler air, may vary depending on factors such as SST, background winds, and first-mode convective intensity (Figure \ref{fig:clim.Feb.Aug.mtn}). Because each type of density current, including sea breezes and cold pools, propagates within a characteristic speed range, the timings associated with these steps largely depend on \textit{ridge–coast distances} (cf. Figure \ref{fig:clim.Hov.coast}b–c) and can differ substantially across islands of varying size. Despite these variations, similar jumps accompanied by discontinuous convective propagation have also been observed in radar measurements on the western coast of Sumatra (see Yokoi et al., 2019, Figs.~5–6).

\subsection{Cold pool regimes}
\label{sec: cold pool regimes}

Cold pools play a critical role in each stage of offshore propagation (Figures \ref{fig:snap.Tz2}-\ref{fig:cross.LBP}). Based on their location, we categorize cold pools into three regimes:

\begin{enumerate}
    \item \textbf{Residual cold pools over land} originate from the first convective mode over sloped terrain to form by sunset and maturing into circular outflows by midnight (Figure \ref{fig:snap.Tz2}). Often triggered approximately 100 km onshore at the sea‐breeze convergence zone, they persist $\sim$10 hours until early morning (Figure \ref{fig:clim.Hov.coast}), potentially aided by cloud–radiative interactions sustaining nocturnal convection \cite{Najarian2025}. These cold pools reach depths of $\sim$750 m (Figure \ref{fig:cross.LBP}) and drive the sharp pre‐sunset coastal wind reversal (Figure~\ref{fig:Hov.Ywind.surface}) and can even penetrate across the coastline into the ocean, most notably along the southwestern shore (Figure~\ref{fig:snap.Tz2}c), making them key components of the sea breeze–cold pool–land breeze interference (as illustrated in Figure~\ref{fig:moistBL.schematic}a–c).

    \item \textbf{Embedded cold pools} form along the advancing gust fronts of sea–land breezes, driven by relatively weak convective cells that propagate with the fronts. Their limited precipitation intensity and lack of a fixed position prevent them from developing into large, long-lived cold pools as seen in other regimes. As they move, these shallow and transient cold pools often merge with the gust fronts, locally enhancing gust wind speeds (Figures~\ref{fig:Hov.Ywind.surface}, \ref{fig:snap.Tz2}, and~\ref{fig:snap.Q2}). However, if the land breeze weakens after sunrise or gradually slows down, while convection becomes more organized or moves faster than the breeze front, the resulting cold pools may extend beyond the breeze boundary, detach from it, and evolve into the next regime.

    \item \textbf{Cold pools over the ocean} form directly over warm SST when offshore convection produces heavy precipitation (Figure \ref{fig:Hov.PR-GPM_C1_SST}). Once land breeze-driven outflows fade, these cold pools sustain and help to organize convection beyond 600 km offshore, spawning successive cells into the next diurnal cycle and contributing to \textit{$\sim$600 km long-distance} offshore propagation (Figures \ref{fig:snap.Q2}, \ref{fig:cross.CPP}). 
\end{enumerate}

In this case, organized offshore convection persists for approximately two diurnal cycles, and the propagation distance is roughly twice that of a typical single-day event. Individual cold pools remain relatively small and expand nearly isotropically. As such, isotropic spreading alone does not inherently favor persistent, offshore-directed propagation. Sustained propagation instead arises from the combined mechanical and thermodynamic impacts of cold pools (third regime) on the boundary layer. Beyond their direct lifting at the gust front, cold pools leave behind persistent post-convective cooling, creating a thermally asymmetric boundary-layer structure that systematically favors new convective initiation on the offshore side of existing cold pools (Figure \ref{fig:cross.CPP}). This thermodynamic memory enables successive regeneration and produces the leapfrogging behavior. Notably, this mechanism remains effective even in February over the northeastern offshore waters, when cross-equatorial monsoonal flow slightly above the boundary layer imposes low-level wind shear that is dynamically unfavorable for offshore cell development. However, this successive offshore convective regeneration is not indefinite, as boundary-layer recovery, diurnal reversal of coastal circulations, and evolving large-scale conditions progressively weaken the asymmetry required for offshore re-initiation.


The \textit{hybrid land breeze} integrates different regimes of cold pools; therefore, the relative contributions of dry pressure-gradient dynamics and moist thermodynamic processes are hypothesized to evolve through different stages (as illustrated in Figure~\ref{fig:monsoon.schematic}c–d). Over land, the residual convective cold pools (first mode / first regime) generated by evaporative cooling, together with the pure land breeze formed by radiative cooling, both act as sources of dense air (see Section \ref{definitions} for details) and merge to form the hybrid land breeze (Figure \ref{fig:cross.LBP}). As this hybrid land breeze propagates offshore, the warm ocean surface promotes the development of \textit{moist patches} within its front (Figure \ref{fig:snap.Q2}). Over the ocean, moist convection (second mode) can increase water vapor through rain evaporation and through surface moisture fluxes enhanced by cold pools (second or third regime), strengthening and curving the moist patches (as illustrated in Figure~\ref{fig:moistBL.schematic}c–d).

The relationship between the fronts of the first-regime cold pools and the hybrid land breeze likely depends on island size or ridge–coast distance (e.g., cf. Figure \ref{fig:clim.Hov.coast} b-c), convective intensity (first mode), and background wind advection. For instance, across many islands of the Maritime Continent, the distance between the mountain ridge and the coastline is considerably shorter than in New Guinea. If the land convection (first mode) is sufficiently strong, the \textit{sea breeze–cold pool–land breeze interference} may complete earlier than the onset of the land breeze driven by radiative cooling. In such cases, the first-regime cold pools may play a more direct, smaller-scale role in the early stages of convective offshore propagation, with their gust fronts extending across the coastline and forming distinct moist patches over the ocean.

\subsection{The cross-equator monsoon}
\label{sec:monsoon}

Precipitation over New Guinea and its surrounding seas is shaped by three main monsoon systems (Figures \ref{fig:clim.Feb.Aug.mtn}b,d): the cross-equatorial monsoon, the Australian Monsoon, and the Asian Monsoon. Land-sea thermal contrasts on the island of New Guinea itself also contribute, almost as a smaller-scale monsoon system. SST anomalies along the southwestern coast (10°S–2°S) show a consistent correlation with those near northern Australia, characterized by much lower SSTs in August, which may inhibit the regeneration of the second mode of convection (Figure \ref{fig:clim.Feb.Aug.mtn}c).

\begin{figure}
 \noindent\includegraphics[width=39pc,angle=0]{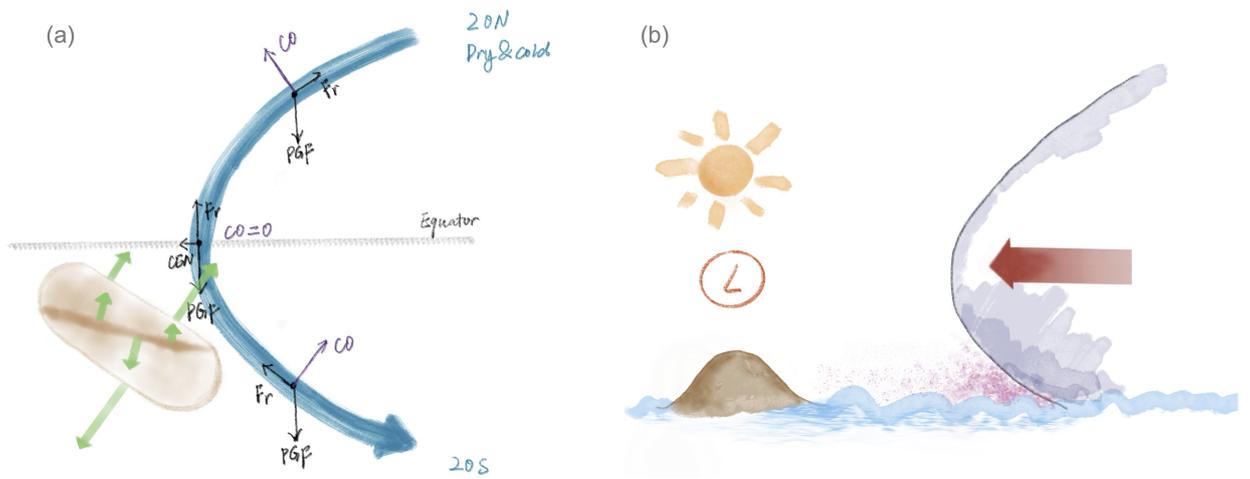}\\
 \caption{Schematic diagram of \textbf{cross-equator monsoonal flow and head} and their roles in diurnal offshore propagation over New Guinea (section \ref{sec:monsoon}): horizontal seasonal dynamic and thermal effects in boreal winter (a), and vertical diurnal thermodynamic effect and moisture convergence by the accumulated and unstable cold and dry monsoon air near the equator (b). In (a), the brown diamond denotes the idealized coastline of New Guinea, and the thick brown line indicates a ridge oblique to it; green arrows mark two distinct convective modes composing the offshore propagation. CEN represents the centrifugal force at the equator, while CO, PGF, and Fr denote the Coriolis, pressure-gradient, and frictional forces forming the three-force balance. In (b), the purple shading represents cold and dry monsoonal air, pink dots indicate water vapor, and the brown hill marks New Guinea with a daytime low-pressure system above. Schematic diagram of the \textbf{hybrid land breeze} over New Guinea at 12~AM (c) and 6~AM (d): dark blue and black curves mark the fronts of the hybrid land breeze and first- and second-regime cold pools (section~\ref{sec: cold pool regimes}), driven by nighttime radiative (light blue) and evaporative (gray) cooling. Red dots indicate water vapor with moist patches near the fronts. Thin orange and brick-red arrows show dense-air mixing and buoyancy-driven motions, while thick orange arrows denote second-mode convection and hybrid land-breeze fronts propagating offshore at 4--7~m~s$^{-1}$.}
 \label{fig:monsoon.schematic}
\end{figure}

\begin{figure}
 \noindent\includegraphics[width=39pc,angle=0]{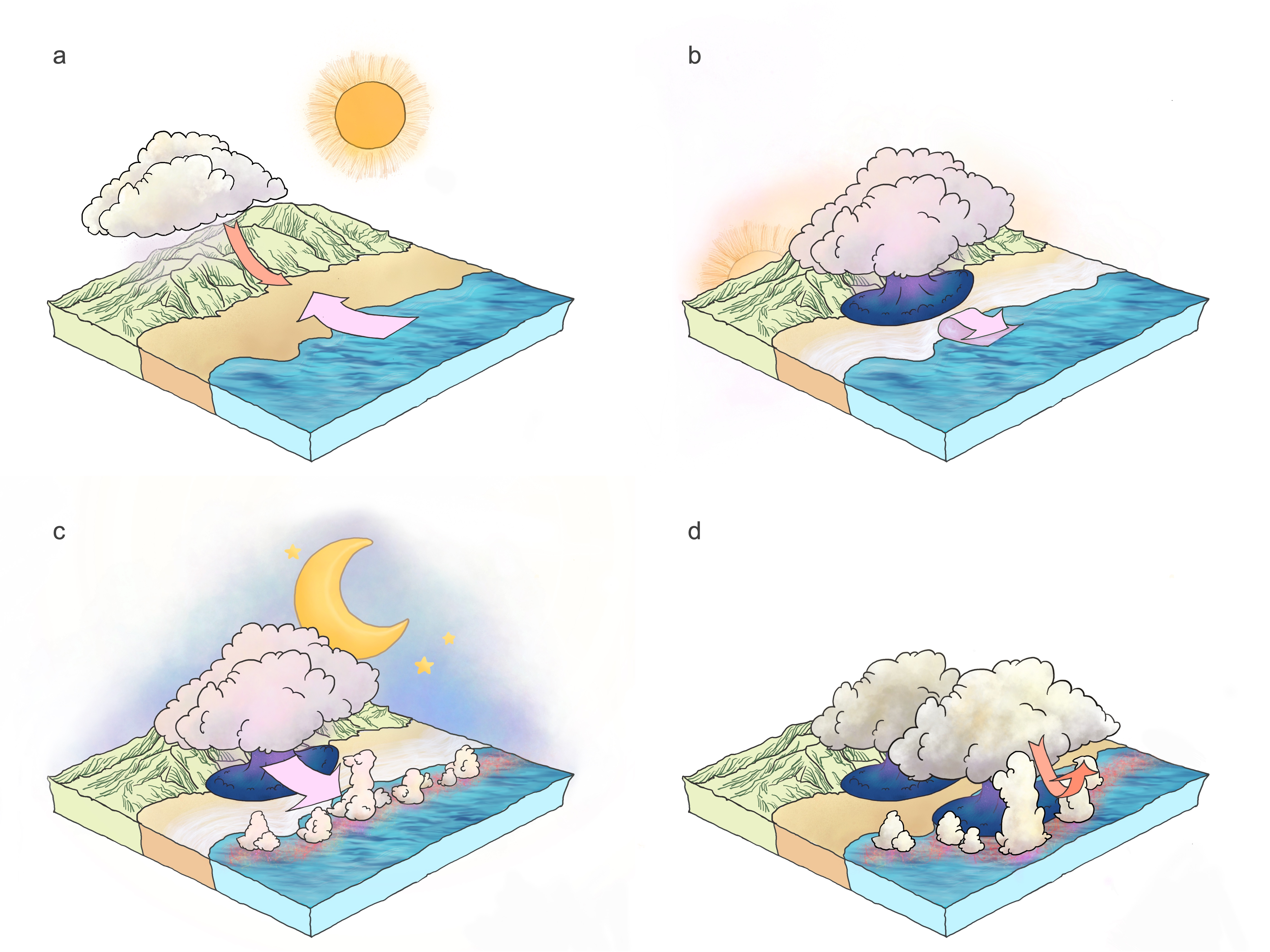}\\
 \caption{Schematic diagram of moist boundary layer dynamics: 
 (a) Daytime heating triggers convection over the ridge in the early afternoon; then convection (first mode) moves fast from the steep terrain slope to the coast when the afternoon sea breeze front continues moving onshore.
(b) First-mode propagation meets a sea breeze front from the warm ocean, enhancing convection and leading to robust cold pools; sea breeze traps cold air in the boundary layer, contributing to further jump; cold pools help the sharp transition from long-lasting sea breeze to earlier initiated observed land breeze before sunset.
(c) Nocturnal land breeze driven by radiative cooling + residual cold pools = hybrid land breeze and new offshore convection (second mode), convection regenerated over the warm ocean; sea breeze–cold pool–land breeze interference (section \ref{concept:SCL interference}); jump between two modes; moist patches (areas with anomalously high water vapor) occur at the land breeze head, parallel to the coastline.
(d) The forward cloud over the ocean grows while the backward cloud over land decays; organized convection (second mode) persists far offshore ($>$200–600 km), even after sunrise the next day; continuing cold pool interactions enhance and curve the moist patches, and enable leapfrogging propagation at intervals of roughly a cold-pool radius.}
 \label{fig:moistBL.schematic}
\end{figure}

During boreal winter, an asymmetry in cooling of the Northern Hemisphere and the Southern Hemisphere results in a cross-equatorial northerly monsoon, a large-scale thermally driven flow. This colder and drier airflow of cross-equatorial monsoons differs from cold pools and land breezes because it occurs higher in the lower troposphere and is more strongly affected by the Coriolis force—an effect that is crucial for sustaining such a large-scale, steady boundary flow on the equatorial $\beta$-plane \cite{wang2006MS}. This effect causes the zonal velocity of the monsoon to change sign as the air crosses the equator and adjusts to reach an Ekman balance with pressure gradient and frictional forces in the lower troposphere (Figure \ref{fig:monsoon.schematic}a). 

Thermodynamically, the cross-equatorial monsoon modifies both SST and boundary-layer instability, shaping the seasonal (Figure \ref{fig:clim.Feb.Aug.mtn}) and meridional (e.g., Figure \ref{fig:clim.Hov.coast}b and c; schematic in Figure \ref{fig:monsoon.schematic}a) variations of diurnal convection. On one hand, the curvature and deceleration of the cross-equatorial flow can lead to cold-air accumulation from the Northern Hemisphere, cooling the boundary layer. On the other hand, acting as a low-level jet, it enhances wind-induced evaporation and some coastal upwelling, thereby decreasing SSTs around the island. Because New Guinea lies south of the equator, both effects may contribute to the disappearance of weaker inland convection associated with the afternoon sea breeze, as well as the second mode on the southwestern flank of the mountain ridge in August (Figure \ref{fig:clim.Feb.Aug.mtn}c).

The cold and dry monsoonal air, with maximum speeds near 1.5 km ($\sim$850 hPa) altitude (see Figures \ref{fig:clim.Feb.Aug.mtn}, S4 and S5 in Supporting Information), interacts with the warm ocean surface, triggering complex thermodynamic processes and vertical mixing that can promote cloud and precipitation. This process may contribute to the third group of precipitation located northwest of New Guinea near the equator (Figure~\ref{fig:2coors}b), which also appears as an organized pattern \cite{tompkins2025}, where the cross-equatorial flow undergoes a complete zonal wind reversal (Figure~\ref{fig:clim.Feb.Aug.mtn}b and Movie S2). In addition, as the flow curves around the equator and its velocity weakens, especially around 500 m ($\sim$950 hPa), wind-driven evaporation from the ocean surface decreases, allowing local SSTs to rise and further destabilize the boundary layer beneath the \textit{monsoonal head} (Figure \ref{fig:monsoon.schematic} b). This enhanced thermodynamic instability occurs alongside the inertial instability associated with large-scale dynamics \cite{wang2006MS} and the centrifugal ejection of faster cold air from the turning flow, thereby increasing the potential for cross-scale interactions.

Monsoonal head also interacts with New Guinea’s diurnal cycle of convection indirectly by synoptic systems and large-scale circulations. For example, Figure \ref{fig:clim.Hov.coast} shows that, in some years, sporadic weak onshore rainfall can persist for a significant amount of time, often converging with offshore propagating signals from the island (e.g., 2004, 2005, 2014, 2018, 2020; see Figure S8 in Supporting Information). Figure \ref{fig:monsoon.schematic}b illustrates the conceptual thermodynamic structure of the monsoonal head and its cross-scale interaction with the island diurnal cycle. During daytime, radiative heating over land and subsequent convection generate a thermal low, which not only strengthens the sea-breeze head but can also attract the nearby monsoonal head toward the island.The dynamic core peaks above the boundary layer ($\sim$1.5 km) and the monsoonal head establishes a zone of enhanced \textit{moisture convergence}, driving humid air to move downward and forward. Unlike moist patches in density currents, the deep monsoonal inflow does not affect surface heat fluxes as much (Figure \ref{fig:snap.LH}), yet interacts strongly with density currents and convection over the ocean (Figures \ref{fig:cross.CPP}, \ref{fig:cross.LBP}), reinforcing and organizing offshore propagation.
Unlike dynamic convergence, moisture convergence imposes a much weaker requirement on vertical structural coherence and instead emphasizes the cumulative effect of moisture supplied by the flow.

Monsoonal flow is a synoptic-scale phenomenon sensitive to its proximity to the island and the dryness and temperature of the air. When it is too close and excessively dry or cold, it suppresses the initiation of the first mode and its offshore propagation. If too distant, its moisture convergence and interaction with the second mode become less effective. Thus, moderate monsoonal distance and intensity are optimal for sustaining both convective initiation and long-distance offshore propagation. This is evidenced by the long offshore propagation distances observed in our case study. These findings align with the work of \citeA{XC2024GWmonsoon}, which shows the role of moderate monsoonal wind speeds in facilitating long offshore propagation through gravity waves.

Beyond monsoons, intraseasonal and interannual variability from the MJO and ENSO further modulate the convective offshore propagation in February (Figures S9–S10, Supporting Information). Active MJO convection over the Maritime Continent amplifies and extends offshore propagation symmetrically from both sides of the mountain ridge (Figure S10), likely through enhanced moisture and upward motion \cite{sakaeda2017, vincent2016NGmjo, vincent2017}; and from the northeastern coastline of New Guinea (Figure S9), probably through westerly wind anomalies during certain MJO phases \cite{wang2006MS}. El Niño winters (e.g., Feb 2010, 2020) display shorter propagation from the northeastern coastline (Figure S9), likely under cooler SSTs accompanied by anomalous anticyclonic circulations north of New Guinea (Figure S1) \cite{wang2000,wang2002}, particularly in the \textit{eastern} Maritime Continent, consistent with \cite{chang2004, jiang2018}; whereas La Niña years (e.g., 2008, 2011, 2021) tend to exhibit longer offshore reach (Figure S9).

\section{Conclusions}
\label{sec:conclusions}

This study identifies two distinct modes of diurnal convection near New Guinea: a “ridge-to-coast” first mode triggered by afternoon heating over high terrain, and an “over-ocean” second mode that re-forms at the coastline after sunset. A novel aspect is that we explicitly separate these two modes, linking the first to the afternoon peak over land and the second to the morning peak over ocean, and clarifying their distinct roles in process-level offshore propagation. These two modes are separated by a pronounced spatial gap, where convection from the first mode dissipates inland, and the second mode abruptly initiates at the shoreline before moving offshore. This explains the commonly observed jump in precipitation signals in both long-term climatological analyses and specific case studies. The convective evolution involves an initial sea breeze and ridge convection, generating low-level convergence and stationary convection. Subsequently, a land breeze reverses the sea breeze and initiates offshore propagation, which is further intensified near the coast by convection and associated cold pool outflows.

Land-sea breezes, convectively generated cold pools, and cross-equatorial monsoonal flows together shape the boundary layer dynamics near the coast. In the afternoon, persistent sea breeze fronts carry cooler air onshore, suppressing or weakening the first mode of convection before it can reach the coast. After sunset, the nocturnal land breeze interacts with residual cold pools, forming a robust \textit{sea breeze–cold pool–land breeze interference}. This process creates moist patches in the marine boundary layer and fuels the second mode of convection over the warm ocean. 

These nighttime processes explain how offshore convection can extend 200–600 km from the coast and persist through morning. While gravity waves have long been considered a critical driver of far-reaching (over 100 km) offshore propagation in the tropics, our results highlight the role of cold pools, the proposed hybrid land breezes, and their moist patches as forcing mechanisms. By interacting with one another and with additional density currents, including the convective system itself, cold pools can maintain and organize storm systems, enabling leapfrogging offshore convection that can reach distances of 600 km or more. Even typically weaker land breezes can lengthen the reach of offshore convection beyond 200 km when they interact with warm ocean surfaces and mix with cold pools.

A modest increase in sea surface temperature substantially enhances the intensity of nighttime convective events, particularly the regeneration of the “over-ocean” mode, and boosts offshore propagation distances. Cross-equatorial monsoons bring cooler, sometimes drier, low-level air, adding variability and sensitivity to these processes. Moderate monsoonal flow and head can further organize and extend offshore precipitation through moisture convergence and interactions with gust fronts from cold pools or land breezes. The coastline perpendicular shallow density current is essential to the diurnal excitation and propagation of the second mode over the warm ocean, while the monsoonal flow modulates its extension and persistence. These combined thermodynamic and dynamic effects introduce regional and seasonal variations in offshore propagation, influencing diurnal precipitation patterns over New Guinea.

By highlighting the multi-scale thermally driven flows fundamental to diurnal convection, this study aims to offer a new perspective on how they interact, couple with moist-patches thermodynamics, and influence the organization and propagation of tropical convection. Future work will quantify the relative contributions of gravity waves and density currents under a range of synoptic and large-scale conditions and assess how the atmospheric and oceanic boundary layers interact under varying SSTs and wind regimes. Such efforts will continue to deepen our understanding of how cross-scale processes shape diurnal convective events, potentially leading to improvements in rainfall forecasts over the Maritime Continent.

%
%

\section*{Conflict of Interest Statement}
The authors have no conflicts of interest to disclose.

\section*{Open Research section}





Data and analysis scripts supporting this study are available at \citeA{Tang2025Data} (\url{https://doi.org/10.5281/zenodo.15493261}). A recorded seminar presenting this work is available online at 
\url{https://www.youtube.com/watch?v=6tlKYKa3GSI}.

\acknowledgments
We acknowledge the funding support from the NSF NCAR Advanced Study Program (ASP) Graduate Visitor Program (GVP) and NOAA grant NA22OAR4310614. The simulations were primarily performed on the Derecho supercomputer provided by the NSF NCAR. We are particularly grateful to Margaret A. LeMone for her contributions during the initial stage of this work. We are especially grateful to Bin Wang, Yi-Leng Chen, and Brian Mapes for their insightful suggestions and inspiration. We also thank Wei Wang for technical assistance. Additionally, discussions with Chidong Zhang,  Richard Rotunno, Yuqing Wang, Fei-Fei Jin, Leishan Jiang, Hyodae Seo, and Pao-Shin Chu have been invaluable in shaping the multi-scale aspects of this study. We thank three anonymous reviewers for their valuable suggestions. M.T. acknowledges support from the ICTP during the revision of this manuscript. This is SOEST contribution number 12146. 



\end{document}